\documentclass[]{raa}        

\usepackage{natbib}
\usepackage[backref]{hyperref}
\hypersetup{
    colorlinks=true,
    linkcolor=blue,
    filecolor=magenta,      
    urlcolor=blue,
    citecolor=blue,
    pdfpagemode=FullScreen,
    }

\usepackage{amsmath}
\usepackage{amssymb}
\usepackage{graphicx}
\usepackage{physics}

\usepackage{xcolor}
\definecolor{light_red}{RGB}{217, 83, 79}
\definecolor{light_orange}{RGB}{255, 173, 96}
\definecolor{light_blue}{RGB}{30, 125, 210}
\definecolor{blue2}{RGB}{146, 53, 184}

\usepackage{soul}
\usepackage[utf8]{inputenc}
\setstcolor{red}
\setul{1pt}{1pt}

\newcommand{\red}[1]{\textcolor{light_red}{#1}}
\newcommand{\blue}[1]{\textcolor{blue2}{#1}}

\newcommand{\m}{\ensuremath{\,{\rm m}}}

\newcommand{\Mpc}{\ensuremath{\,{\rm Mpc}}}
\newcommand{\K}{\ensuremath{\, {\rm K}}}

\newcommand{\rad}{\ensuremath{\,{\rm rad}}}
\newcommand{\sr}{\ensuremath{\,{\rm sr}}}

\begin{document}

\title{A simulation of calibration and map-making errors of the Tianlai cylinder pathfinder array}
\titlerunning{Tianlai Cylinder Simulation}

\volnopage{Vol.0 (20xx) No.0, 000--000}      
\setcounter{page}{1}          

    \author{Kaifeng Yu \inst{1,2,3}
    \and Fengquan Wu \inst{1,\blue{2}}
    \and Shifan Zuo \inst{1,\blue{2}}
    \and Jixia Li \inst{1,\blue{2}}
    \and Shijie Sun \inst{1,\blue{2}}
    \and Yougang Wang \inst{1,\blue{2}}    
    \and Xuelei Chen$^*$
        \inst{1,2,3,4,5}
    }
\institute{National Astronomical Observatories, Chinese Academy of Sciences, Beijing 100101, 
 China; \\
\and
Key Laboratory of Radio Astronomy and Technology, Chinese Academy of Sciences, Beijing 100101, China;\\
\and
School of Astronomy and Space Science, University of Chinese Academy of Sciences, Beijing 100049, China\\
\and 
Key Laboratory of Cosmology and Astrophysics (Liaoning) \& College of Sciences, Northeastern University, Shenyang 110819, China\\
\and
Center of High Energy Physics, Peking University, Beijing 100871, China. \\
$^*${\it xuelei@cosmology.bao.ac.cn}
}

\abstract{
The Tianlai cylinder array is a pathfinder for developing and testing 21cm intensity mapping techniques. In this paper, we use numerical simulation to assess how its measurement is affected by thermal noise and the errors in calibration and map-making process, and the error in the sky map reconstructed from a drift scan survey. Here we consider only the single frequency, unpolarized case. The beam is modelled by fitting to the electromagnetic simulation of the antenna, and the variations of the complex gains of the array elements are modelled by Gaussian processes. Mock visibility data is generated and run through our data processing pipeline. We find that the accuracy of the current calibration is limited primarily by the absolute calibration, where the error comes mainly from the approximation of a single dominating point source. We then studied the $m$-mode map-making with the help of Moore-Penrose inverse. We find that discarding modes with singular values smaller than a threshold could generate visible artifacts in the map.  The impacts of the residue variation of the complex gain and thermal noise are also investigated. The thermal noise in the map varies with latitude, being minimum at the latitude passing through the zenith of the telescope. The angular power spectrum of the reconstructed map show that the current Tianlai cylinder pathfinder, which has a shorter maximum baseline length in the North-South direction, can measure modes up to $l \lesssim 2\pi b_{\rm NS}/\lambda \sim 200$ very well, but would lose a significant fraction of higher angular modes when noise is present. These results help us to identify the main limiting factors in our current array configuration and data analysis procedure, and suggest that the performance can be improved by reconfiguration of the array feed positions.   
}
\maketitle



\section{ Introduction}
\label{sec:introduction}

Neutral hydrogen (HI) is ubiquitous in our Universe, it provides a way to probe the early Universe, and can serve as a tracer of the large-scale matter distribution to reconstruct the expansion history of the Universe (e.g. \citealt{2006PhR...433..181F, 2012RPPh...75h6901P}).  Making use of the baryon acoustic oscillation (BAO) which can be treated as a cosmological standard ruler, we can measure the dynamics of dark energy. In the past, the BAO has been measured in the optical galaxy redshift survey(e.g.\citealt{2003ApJ...594..665B, 2003ApJ...598..720S, 2005ApJ...633..560E}).
Radio observations would complement the optical surveys, and may also be applied to redshifts which so far has not been observed. 
A promising technique called 21 cm intensity mapping (IM) offers a economical and efficient way to probe this feature at our interested scale(e.g. \citealt{2008PhRvL.100i1303C, 2015ApJ...803...21B, 2017MNRAS.466.2736V}). With this technique, instead of detecting individual galaxies, one can measure the aggregate emission from many galaxies in a patch of sky over a range of frequencies (i.e. redshift) to map the three-dimensional structure. This technique has been demonstrated by cross-correlating the data from the Green Bank Telescope (GBT) and the optical galaxy survey \citep{2010Natur.466..463C, 2013ApJ...763L..20M}). Recently, from the MeerKAT survey, another detection of the cross-correlation power spectrum between the HI IM and optical survey has also been reported \citep{2023MNRAS.518.6262C}. Meanwhile, \cite{2023arXiv230111943P} presents a direct detection of HI power spectrum on $\Mpc$ scale. Other existing or ongoing IM experiments focusing on the late-time cosmology include both single-dish telescopes and interferometers, such as FAST\citep{2020MNRAS.493.5854H, 2023arXiv230506405L}, BINGO\citep{2013MNRAS.434.1239B}, CHIME\citep{2022arXiv220201242C}, Tianlai\citep{2012IJMPS..12..256C, 2020SCPMA..6329862L}, and the future next-generation radio telescope SKA\citep{2015aska.confE..19S} also can be used for HI IM observation. The future HI IM surveys will provide an unprecedentedly larger volume of the observable Universe than ever.

However, the 21 cm cosmology also faces great challenges including the contamination from the foreground which is 4-5 orders of magnitude brighter than the cosmological signal, and the complexity in data analysis\citep{2020PASP..132f2001L}. Although in principle the foreground radiation has a smooth spectrum while the 21cm signal a stochastic one, in practice, the observed spectrum of the  foreground is not smooth, which is distorted by instrumental effects, such as the chromatic response of beam and the polarization leakage, which mix the spectral and spatial modes respectively (e.g. \citealt{2009ApJ...695..183B, 2021MNRAS.504..208C}). 
The main strategies for dealing with foreground include subtraction, avoidance, and suppression\citep{2016MNRAS.458.2928C,2018ApJ...864..131K}. A number of foreground subtraction techniques have been developed, e.g.  polynomial fitting\citep{2006ApJ...650..529W}, Principal Component Analysis (PCA)\citep{2012MNRAS.419.3491L}, and Gaussian Process Regression (GPR)\citep{2018MNRAS.478.3640M}.

The instrument response of the telescope is itself determined by calibration measurements. As any physical measurement, the calibration measurement also has its own measurement error, this error will propagate in the data processing and induce error in the final result. As the foreground is much stronger than the 21cm signal, the error induced by the instrument calibration could significantly affect the final result. One way to assess the impact of calibration error is to conduct an end-to-end simulation: various instrumental effects are modeled in the simulation to mimic the actual observation, the mock data generated this way is then processed with the same data processing pipeline, in order to check the data analysis algorithms and evaluate the effects of the calibration error on the final results.

The Tianlai project is an experiment aimed at exploring the hardware design and data analysis technique for 21 cm intensity mapping, it includes a cylinder array and a dish array. The basic performance of the Tianlai array has been analyzed in \citet{2020SCPMA..6329862L} for the cylinder pathfinder, and in \citet{2021MNRAS.506.3455W} for the dish pathfinder. In this paper, we investigate  the impact of calibration error on the map-making of the Tianlai cylinder pathfinder array. This will also be useful in the interpretation of the observational data, and testing the validity and efficiency of algorithms in the data processing pipeline. 

\begin{figure}
    \centering
    \includegraphics[width=0.6\textwidth]{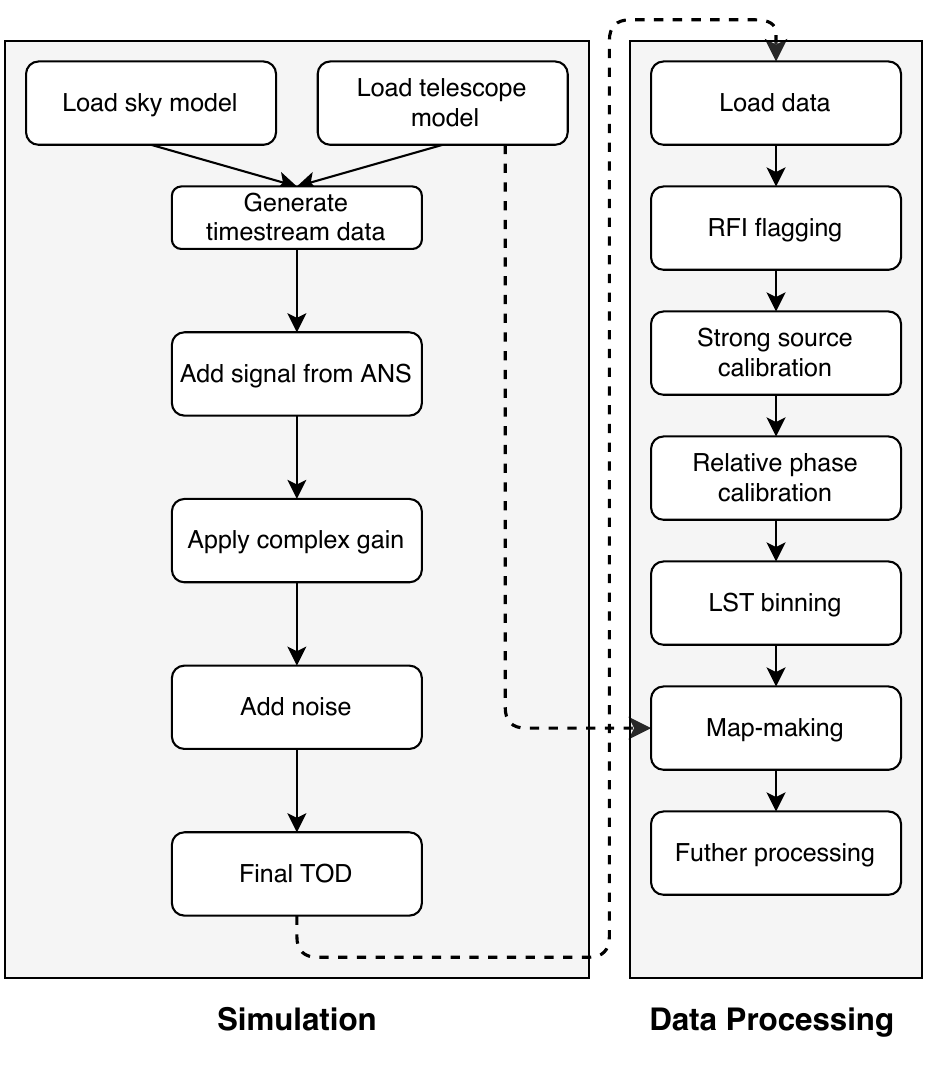}
    \caption{The flowchart of main steps in simulation and data processing of \texttt{tlpipe}.}
    \label{fig:flowchart}
\end{figure}

The complex gains of the array elements are calibrated by observing bright radio sources in the sky (absolute calibration), and a periodically broadcasting noise source (relative calibration). 
In this paper we mainly consider the errors in these calibrations, and how these errors would affect the synthetic maps made from the observation, using the $m$-mode analysis method. The main steps of simulation are shown in the left panel of Fig.~\ref{fig:flowchart}. We use a sky model to generate the mock interferometric visibility data, and then process it with the Tianlai data processing pipeline (\texttt{tlpipe})\citep{ZUO2021100439}, to produce the calibrated data and sky map. We compare these with the input, to determine the error in the calibration, map-making, and power spectrum estimation. 

The paper is organized as follows. We first describe the basic models which are used in our simulation, including the telescope beam, the sky model, the receiver noise, and gain variation in Section \ref{sec:simulation}; then we describe the calibration procedure and apply the data processing pipeline to the mock data for calibration in Section \ref{sec:tlpipe}. We then describe the $m$-mode formalism of map-making and the results in Section \ref{sec:map-making}. Finally, we conclude in Section \ref{sec:conclusion}.

\section{Simulation}
\label{sec:simulation}

For radio interferometers, the correlation of  voltages between a pair of antenna receivers, or  \textit{visibility}, is related to the sky temperature as  
\begin{eqnarray}
    V_{ij}(t) &=& g_i(t)g_j^{*}(t) \int T(\vb*{\hat{n}}) A_{i}(\vb*{\hat{n}}; t) A_{j}^{*}(\vb*{\hat{n}}; t) e^{2\pi i \vb*{\hat{n}} \cdot \vb*{u}_{ij}(t)} \dd^{2}\vb*{\hat{n}}+n_{ij}(t)\\ \label{eq:vis_eq}
    &=& g_i(t)g_j^{*}(t)\int d^{2} \vb*{\hat{n}} B_{ij}(\vb*{\hat{n}}; t) T(\vb*{\hat{n}}) + n_{ij}(t), 
    \label{eq:vis_eq1}
\end{eqnarray}
where $\vb*{\hat{n}}$ denotes the direction on the sky, $T(\vb*{\hat{n}})$ is the brightness temperature in that direction.
$\vb*{u}_{ij} = (\vb*{r}_i - \vb*{r}_j) / \lambda$ is the baseline vector between the two 
feeds $i$ and $j$ in unit of the wavelength, $A_{i}(\vb*{\hat{n}})$ is the primary beam of feed $i$, $g_i(t)$ is the complex gain of feed $i$ to be calibrated, and $n_{ij}(t)$ is the receiver noise, which we assume to have zero mean.   
The beam transfer function $B_{ij}(\vb*{\hat{n}}; t)$ is 
\begin{equation}
    B_{ij}(\vb*{\hat{n}}; t)= A_{i}(\vb*{\hat{n}}; t) A_{j}^{*}(\vb*{\hat{n}}; t) e^{2\pi i \vb*{\hat{n}} \cdot \vb*{u}_{ij}(t)}.
    \label{eq:btf}
\end{equation}
For simplicity, we shall assume that all 
$A_{i}(\vb*{\hat{n}}; t)$  are the same in this paper. 
For the Tianlai cylinder which is fixed on ground, the rotation of Earth generates varying visibility.

\subsection{Telescope Configuration and Beam Model}\label{beam-model}

The Tianlai cylinder array pathfinder has 3 cylindrical reflectors with 15m width, and there are 31, 32, 33 feeds spaced along the focal axis respectively, which is in the North-South direction, the longest distance between feeds on each cylinder is 12.4m, though the cylindrical reflector itself is 40 m long. The purpose of the unequal spacing is to reduce the grating lobes, an effect appears when the minimum feed spacing is greater than half observational wavelength \citep{2016RAA....16..158Z}. The adopted simulation parameters are listed in Table \ref{tbl:table1}.

\begin{table}
    \centering
    \caption{Simulation parameters.}
    \begin{tabular}{cc}
        \hline
        \hline
        Parameters & Value\\
        \hline
        Latitude		 & $44.15^\circ$ N \\
        Cylinder numbers & 3\\
        Cylinder width   & 15 m \\
        Feeds			 & 31,32,33 [from East to West] \\
        Longest NS Length & 12.4 m \\
        System Temperature	& 90 K\\
        Bandwidth 		 &  700-800 MHz\\
        Channel width    &   122 kHz \\
        \hline
    \end{tabular}
		\label{tbl:table1}
\end{table}

The primary beam pattern of the Tianlai cylinder pathfinder has not been measured, especially along the North-South direction, though along the East-West direction it can be derived by observing the transit of a strong radio source \citep{2019AJ....157...34Z, 2020SCPMA..6329862L}. 
Here we model the primary beam as a product of a Gaussian function along the North-South direction and a Fraunhofer diffraction solution along the East-West, 
$$
A(\hat{\vb*{n}}) = A_{\text{NS}} (\sin^{-1}(\hat{\vb*{n}} \cdot \hat{\vb*{x}}); \theta_{\text{NS}}) \times A_{\text{EW}} (\sin^{-1}(\hat{\vb*{n}} \cdot \hat{\vb*{y}}), F, \theta_{\text{EW}}) 
$$
where $\hat{\vb*{x}}$ and $\hat{\vb*{y}}$ are the unit vector pointing East and North, respectively, and $F$ is the focal ratio.
In the North-South direction, the beam amplitude has the form
\begin{equation}
A_{\text{NS}} (\theta; \theta_{\text{NS}}) = \exp \left[ - 4 \ln2 \left( \frac{\theta}{\theta_{\text{NS}}} \right)^2 \right]
\label{eq:beam-ns}
\end{equation}
where $\theta_{\text{NS}}(\nu) = \alpha \frac{\lambda}{D_{\text{NS}}}$ and $D_{\text{NS}} = 0.3$m as the size of the Tianlai cylinder feeds. For the beam in the East-West direction, we take the form
\begin{align}
    \nonumber
     A_{\text{EW}}(\theta;F, \theta_{\text{EW}}) 
     &\propto 
     \int^{\frac{W}{2}}_{-\frac{W}{2}}A_D\left(2 \tan^{-1}\left(\frac{x}{2FW}\right) ;\theta_\text{EW}\right) e^{-ikx\sin\theta} \dd x \\
     &\propto \int^{\frac{1}{4F}}_{-\frac{1}{4F}} A_D\left(2 \tan^{-1} u ;\theta_\text{EW}\right) e^{- i\frac{4\pi F W u}{\lambda} \sin \theta} \dd u,
    \label{eq:beam-ew}
\end{align}
where $W$ is the width of the cylinder, and $A_D$ is taken as
\begin{equation*}
    A_{D}(\theta; \theta_{\text{FWHM}}) = \exp \left[ -\frac{\ln 2}{2} \frac{\tan^2\theta}{\tan^2(\theta_{\text{FWHM}}/2)} \right],
\end{equation*}
which is consistent with \cite{2015PhRvD..91h3514S}. We choose the parameter $\alpha$ in $\theta_{\text{NS}}$ and $F$, $\theta_{\text{EW}}$ in $A_{\text{EW}}$ by fitting the result of the electromagnetic(EM) field simulation \citep{2022RAA....22f5020S}.

\begin{figure}
    \centering
    \includegraphics[width=0.8\textwidth]{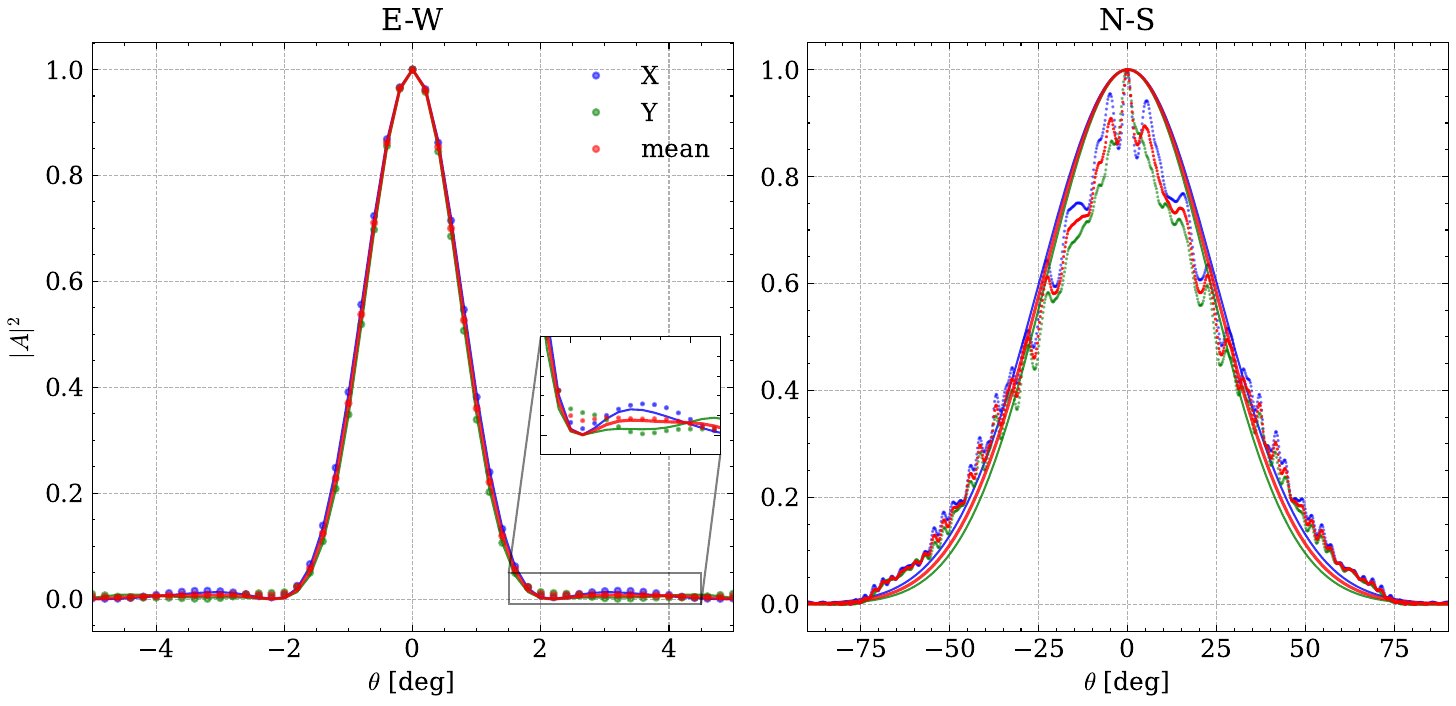}
    \caption{The EM simulated normalized power gain(dot) of the X(blue) and Y(green) polarization feed at 750 MHz and the corresponding fitted beam in the East-West direction(\textit{left}) and North-South direction(\textit{right}). The red line fits the average of the two polarization, which is used for our unpolarized case.}
    \label{fig:beam-fitting}
\end{figure}

In Fig. \ref{fig:beam-fitting}, we show the fitted power beam $|A|^2$ with the EM simulated normalized gain at $750\text{MHz}$ in the East-West direction and the North-South direction separately. We fit the data with the mean of the X and Y polarization for our unpolarized case, which gives the fitted parameters $\alpha = 1.04$, $F = 0.2$ and $\theta_{\text{EW}} = 2.74$. This may still differ from the actual case, but for the purpose of simulation it should be sufficient.

\subsection{Sky Model}\label{sec:sky-model}
We generate the sky maps including foreground components and the cosmological 21 cm signal using the publicly available code \texttt{cora}\footnote{\url{https://github.com/radiocosmology/cora}}\citep{2014ApJ...781...57S,2015PhRvD..91h3514S}. 
The primary foreground components consist of diffuse synchrotron emission from the Galaxy and emission from the extragalactic point sources. The base map is generated by extrapolating the 408 MHz map \citep{1982A&AS...47....1H} with spectral index from \cite{2008A&A...490.1093M}, and Gaussian random fluctuations in brightness and spectral index are added to account for small scale fluctuations. The emission of extragalactic point sources involves three components: bright point sources in the   VLSS and NVSS catalogues; synthetic dimmer sources, constructed by drawing from the point source distribution in \citet{2002ApJ...564..576D}; unresolved background of dimmer sources, generated by drawing a Gaussian realization from an angular power spectrum for point sources model. The redshifted 21cm signal is treated as a Gaussian fluctuation, generated according to a given angular power spectrum.

\begin{figure}
    \centering
    \includegraphics[width=0.48\textwidth]{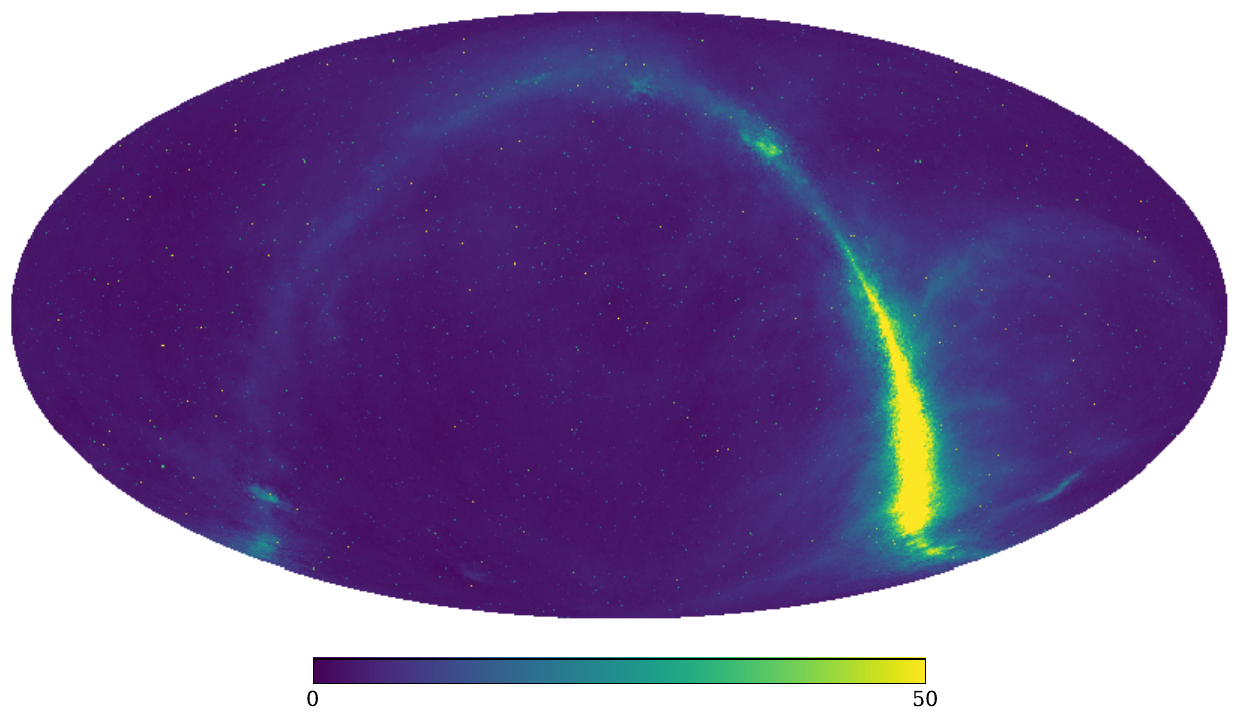}
    \includegraphics[width=0.48\textwidth]{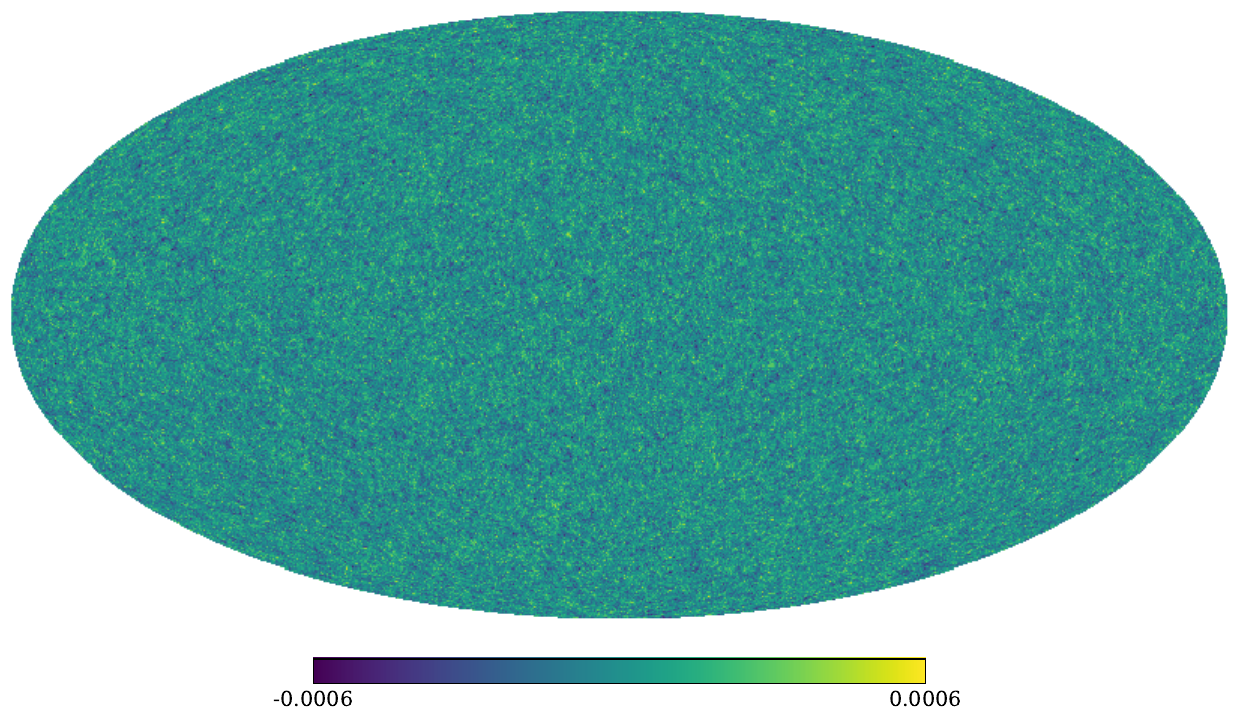}
    \caption{The simulated sky maps at 750 MHz of foreground including diffuse synchrotron emission and extragalactic point sources(\textit{top}) and 21 cm signal(\textit{bottom}).}
    \label{fig:skymap}
\end{figure}

We illustrate the simulated sky map of the foreground and 21 cm signal at 750 MHz in Fig. \ref{fig:skymap}, though in the present study, we will deal mainly with the much stronger foreground radiation. The maps are pixelated using the HEALPix scheme\citep{2005ApJ...622..759G}, with $N_{\text{side}} = 512$ corresponding to an angular resolution of 6.87 arcmin.

\subsection{Receiver Noise}

We assume the noise in the visibility of each feed pair at a given frequency follows a Gaussian distribution, of which the variance at any particular time can be modeled as 
\begin{equation}
    \sigma_{\rm noise} = \frac{\Omega T_{\text{sys}}}{\sqrt{2} \sqrt{ t_{\text{int}} \Delta \nu}}
    \label{eq:noise-sigma}
\end{equation}
where $\Omega = \int |A(\hat{\vb*{n}})|^2 \dd^2 \hat{\vb*{n}}$ is the beam solid angle, $T_{\text{sys}}$ is the system temperature, $t_{\text{int}}$ is the integration time, and $\Delta \nu$ is the bandwidth of frequency channel, the $\sqrt{2}$ in denominator is for an average of the two polarizations--in this work we neglect the difference in the two polarizations. As analyzed in \citet{2020SCPMA..6329862L}, the average system temperature of the Tianlai cylinder pathfinder is 90 K. Currently for the Tianlai cylinder pathfinder $t_{\text{int}} = 4$ s, $\Delta \nu = 122$ kHz. For our simulated beam,  $\Omega \approx 0.034~ {\rm sr}$.  So we get 
$\sigma_{\rm noise}/\Omega=0.091 \K$ per beam, or 
$\sigma_{\rm noise}\approx 0.0031 \K \cdot \sr $  for a one day observation. 
In the sampling of the complex noise data, both the standard deviation in the real and imaginary part are taken as $\sigma_{\text{noise}} / \sqrt{2}$.

\subsection{Complex Gain Fluctuation and Cable Delay}
\begin{figure}
    \centering
    \includegraphics[width=0.98\textwidth]{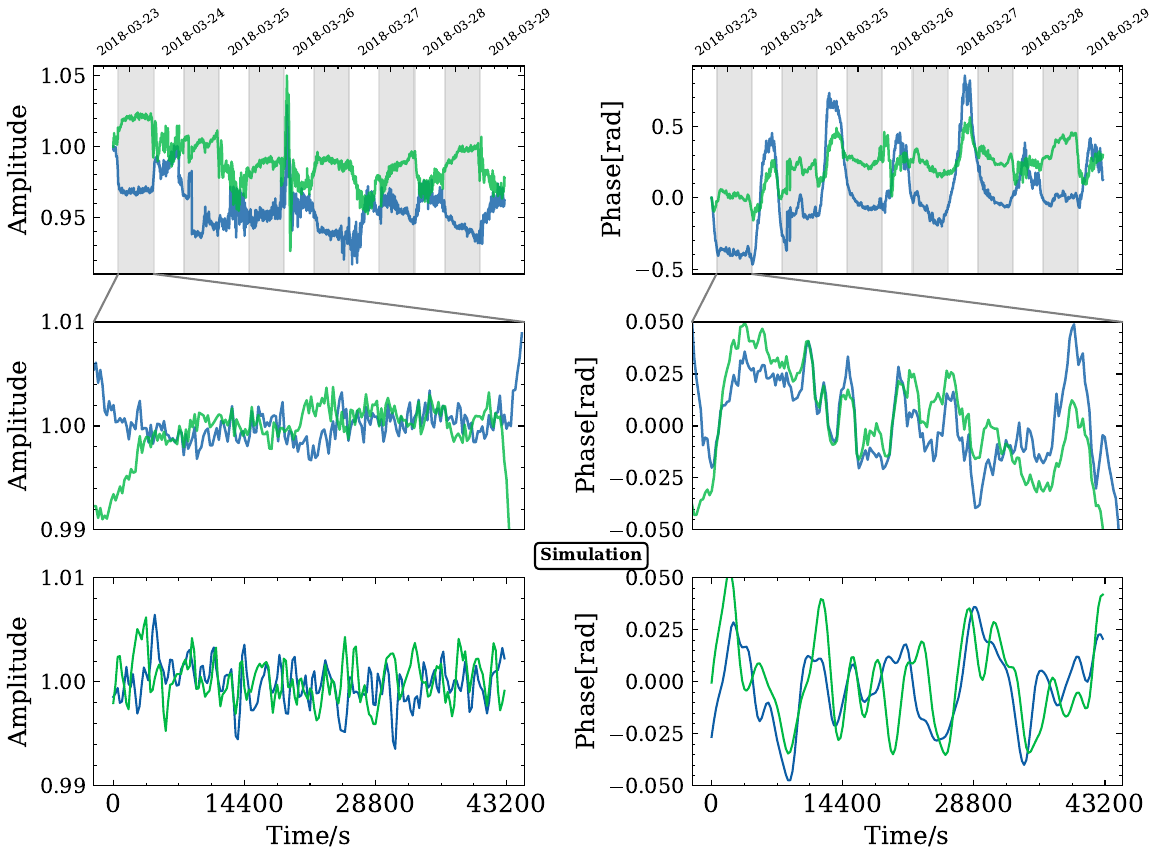}
    \caption{The variation of the relative amplitude(\textit{top left}) and phase(\textit{top right}) of the complex gain for 2 feed channels (plotted in blue and green curves) of the Tianlai cylinder pathfinder in 6 consecutive days starting from 2018/03/22. The shadow area corresponds to the observation in the nighttime(12:00-0:00 UTC). The \textit{middle} panels shows a zoom up of the data of the first night. In the \textit{bottom} panels we show an example of our simulated complex gain variation. }
    \label{fig:gain_example}
\end{figure}

We model the complex gain of each input channel as
\begin{equation}
    g(\nu, t) = |g(\nu,t)|e^{i (2\pi \nu \tau + \varphi(\nu,t))}
    \label{eq:gain_model1}
\end{equation}
where $\tau$ is the instrumental time delay, and 
$\varphi$ is a residue term in the phase which is non-linear with respect to the frequency $\nu$. Due to temperature variation, we found $\tau$ varies at the level of a few to a few tens of picoseconds\citep{2019AJ....157...34Z}, corresponding to millimeter to centimeter changes in the optical cable, and accounts most of the instrument phase variation. The top and middle panels of Fig.\ref{fig:gain_example} show the normalized amplitude and phase of the gain for the 
X-polarization of two feeds on the Tianlai array, which are fairly typical. As we can see from the figure, although there are large variations in the phase during the day time, the phase variation during the night is relatively small and the relative amplitudes are quite stable.

In our simulation, we model the complex gain of one frequency channel as  (we omit the variable $\nu$ here)
\begin{equation*}
    g(t) = (1 + \Delta g(t)) e^{i2\pi \nu\phi(t)}
\end{equation*}
where we have combined the time delay $\tau$ and residue phase $\varphi$ into a single phase $\phi$, as we consider here a single frequency channel. In the actual instrument, a large part of the change in time delay comes from the change of cable length due to temperature variation, so it is not completely random but dependent on the diurnal variation of temperature. However, these changes will be corrected by the calibration procedure, what really matters is the residue for the calibration. So it is sufficient here to model this change as a Gaussian process.  We generate 
the fluctuations in amplitude $\Delta \vb*{g}$ and phase $\Phi$ separately from multivariate normal distributions, \\
$$\Delta \vb*{g} \sim \mathcal{N}\left(\vb*{0}, \vb*{\Sigma}^{\text{amp}}\right), \qquad
\Phi \sim \mathcal{N}\left(\vb*{0}, \vb*{\Sigma}^{\text{phs}}\right).$$ 
The fluctuations should be correlated in time, we model this by adopting 
the covariance function $\vb*{\Sigma}^{\text{amp}}$ and $\vb*{\Sigma}^{\text{phs}}$ as 
\begin{equation}
    \Sigma_{ij} = \sigma^2 \exp(-\frac{(t_i - t_j)^2}{2\xi^2}), 
    \label{eq:corr_func}
\end{equation}
where $\sigma$ and $\xi$ gives the fluctuation amplitude and correlation length respectively.

Inspired by the variation in Fig.\ref{fig:gain_example}, we take $\sigma^{\text{amp}} = 0.002$ for the amplitude part, and $\sigma^{\text{phs}} = 3$ picosecond for the phase part. It is more difficult to get the correlation length, as we do not have sufficiently accurate measurement. Here we take the correlated time to be $\xi^{\text{amp}} = 300$s, and $\xi^{\text{phs}} =  900$s. This model is not derived by a fitting the observation data, but they do give variations comparable in magnitude to the actual case, so that we could have some idea of the effect of variation of the complex gains.

\begin{figure}
    \centering
		\includegraphics[width=0.6\textwidth]{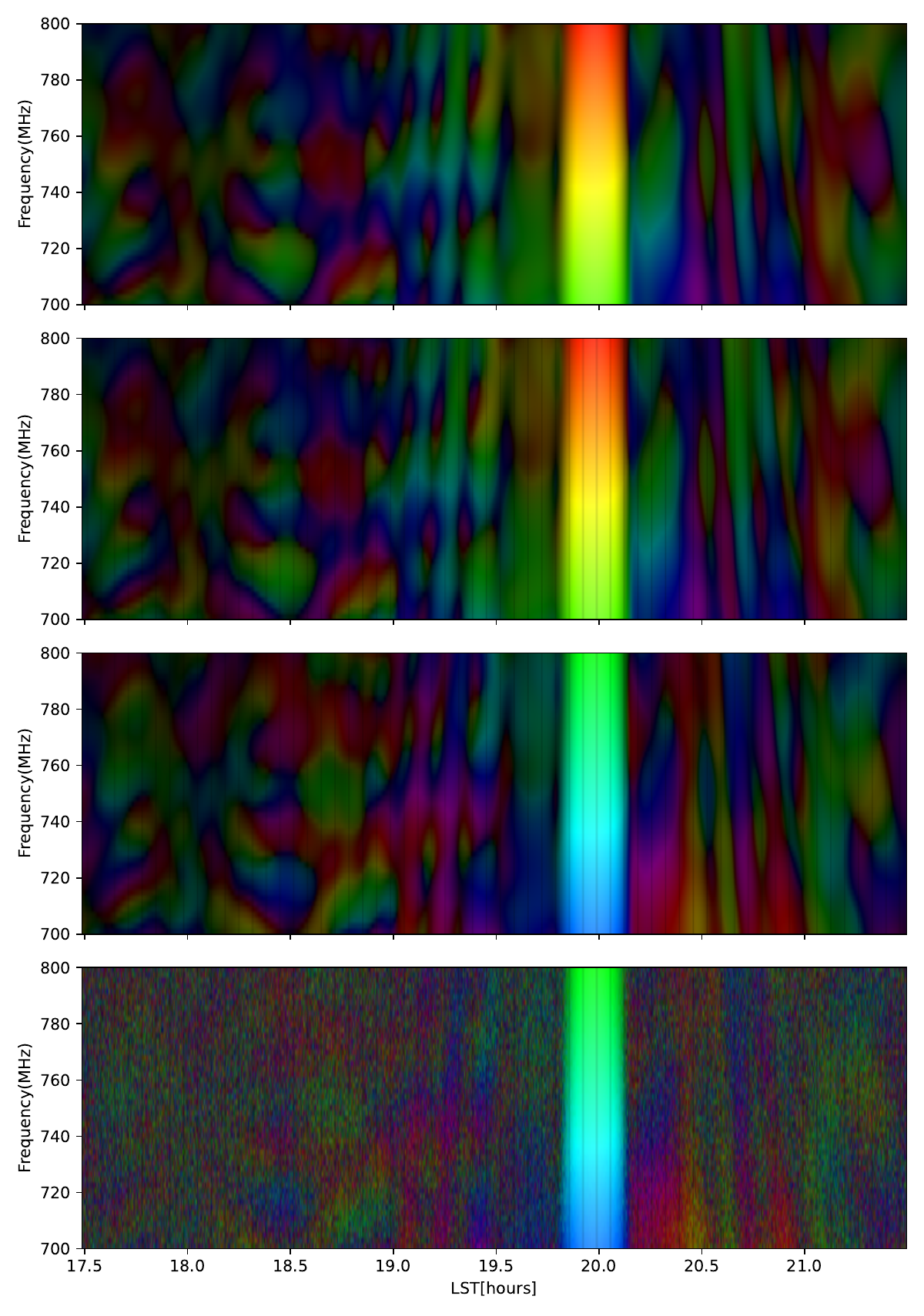}
		\caption{The timestream data of baseline A1-A31 in 4 hours from 700MHz to 800MHz at each main stage of applying instrumental effects. From top to bottom, (1): The underlying visibility; (2): Add signal from noise source (the periodic vertical line); (3): Apply the cable delay; (4): Add Gaussian noise with $N_\text{day} = 1$.}
    \label{fig:tstream_plot}
\end{figure}

The simulated visibility of a duration of 4 hours for a pair of distant feeds on the same cylinder (A1-A31) is shown as a waterfall in Fig. \ref{fig:tstream_plot}, The visualization scheme is from \texttt{SageMath} \footnote{\url{https://doc.sagemath.org/html/en/reference/plotting/sage/plot/complex_plot.html}}, where the brightness and color represents the magnitude and phase of the complex visibility. In the top panel, we show the ideal visibility generated from the sky map. The pattern is produced by the sky model, and the bright feature is the transit of the bright source (Cygnus A) across the primary beam. The color changes with frequency, as we would expect for a baseline of fixed physical length which induce different phase at different wavelength. 
In the second panel from the top, we added the periodically broadcasting noise source (described in the next section), which appear as regularly spaced fine vertical lines. The third panel from the top shows the addition of gain variation. In the bottom panel, we added noise, as would be obtained from a single day observation. Note that in the real data, there would be cross-talks between the different receiver channels, which would induce `correlated noise' which persist over time and show up as horizontal features in the waterfall plot(see \citet{2020SCPMA..6329862L}), here we ignore such noise.

\section{Calibration}
\label{sec:tlpipe}
After generating the mock visibility data using the sky model and telescope model described above, we pass the simulated data to the Tianlai data processing pipeline \texttt{tlpipe}\footnote{\url{https://github.com/TianlaiProject/tlpipe} }\citep{ZUO2021100439}.  Its main tasks include data distribution, RFI flagging, calibration, local sidereal time(LST) binning, map-making, and some other utilities for data selection and analysis. The sketchy data processing procedure from handling raw data to scientific products is shown in the right panel of Fig. \ref{fig:flowchart}. Below we summarize the calibration tasks.

In an interferometer array, the amplitude and instrument phase of the receiver gain $g_{i}$ may vary in time, and the data needs to be calibrated. This can be done by observing known sources. Two kind of calibration sources are used for the Tianlai array:
(1) strong point source on the sky when they are transiting over the fixed Tianlai primary beam, which can provide calibration for both the amplitude and phase of the receiver gain, we shall call this {\bf absolute calibration}; (2) an artificial calibrator noise source (CNS) to calibrate the relative phase change in the complex gain, which we shall call {\bf relative calibration}. 
In \texttt{tlpipe}, the complex gain is calibrated for each frequency channel individually. We will briefly summarize the algorithm here, refer to \citep{2019AJ....157...34Z} for more details.

\subsection{Absolute Calibration}
An absolute calibration can be performed with the help of a bright point source on the sky. When it transits through the main beam of the telescope, we assume the signal from it when transiting dominates the visibility data, which is
\begin{equation}
    V_{ij}^{0} = S_c g_i g_j^{*} A_i(\hat{\vb*{n}}_0) A_j^{*}(\hat{\vb*{n}}_0) e^{i 2\pi \hat{\vb*{n}}_0 \cdot (\vb*{u}_i - \vb*{u}_j)}
    \label{eq:ps_vis}
\end{equation}
where $S_c$ is the flux of the strong point source, $\hat{\vb*{n}}_0$ is its position. Due to RFI or receiver malfunction, some of the data may be corrupted and appear as outliers. In the presence of outliers and noise, the whole data can be written in matrix form as $\vb*{V = V^{0} + S + N}$, where $\vb*{S}$ is a sparse matrix which might come from residual RFI or missing values, $\vb*{N}$ represents the contribution from noise. We can perform a Stable Principal Component Analysis(SPCA) on $\vb*{V}$ to extract $\vb*{V^0}$, then
\begin{equation*}
     V_{ij}^{0} = S_c G_i G_j^{*},
\end{equation*}
where ${G}_i = g_i A_i(\hat{\vb*{n}}_0) e^{-2\pi i \hat{\vb*{n}}_0 \cdot \vb*{u}_i}$ or in matrix-vector form
\begin{equation}
    \vb*{V}^{0} = S_c \vb*{G G^{*}}.
    \label{eq:eigen_cal}
\end{equation}
Note that the vector $\vb*{G}$ is an eigenvector of $\vb*{V}^{0}$ corresponding to its largest eigenvalue. 

\begin{figure}[ht]
    \centering
    \includegraphics[width=0.98\textwidth]{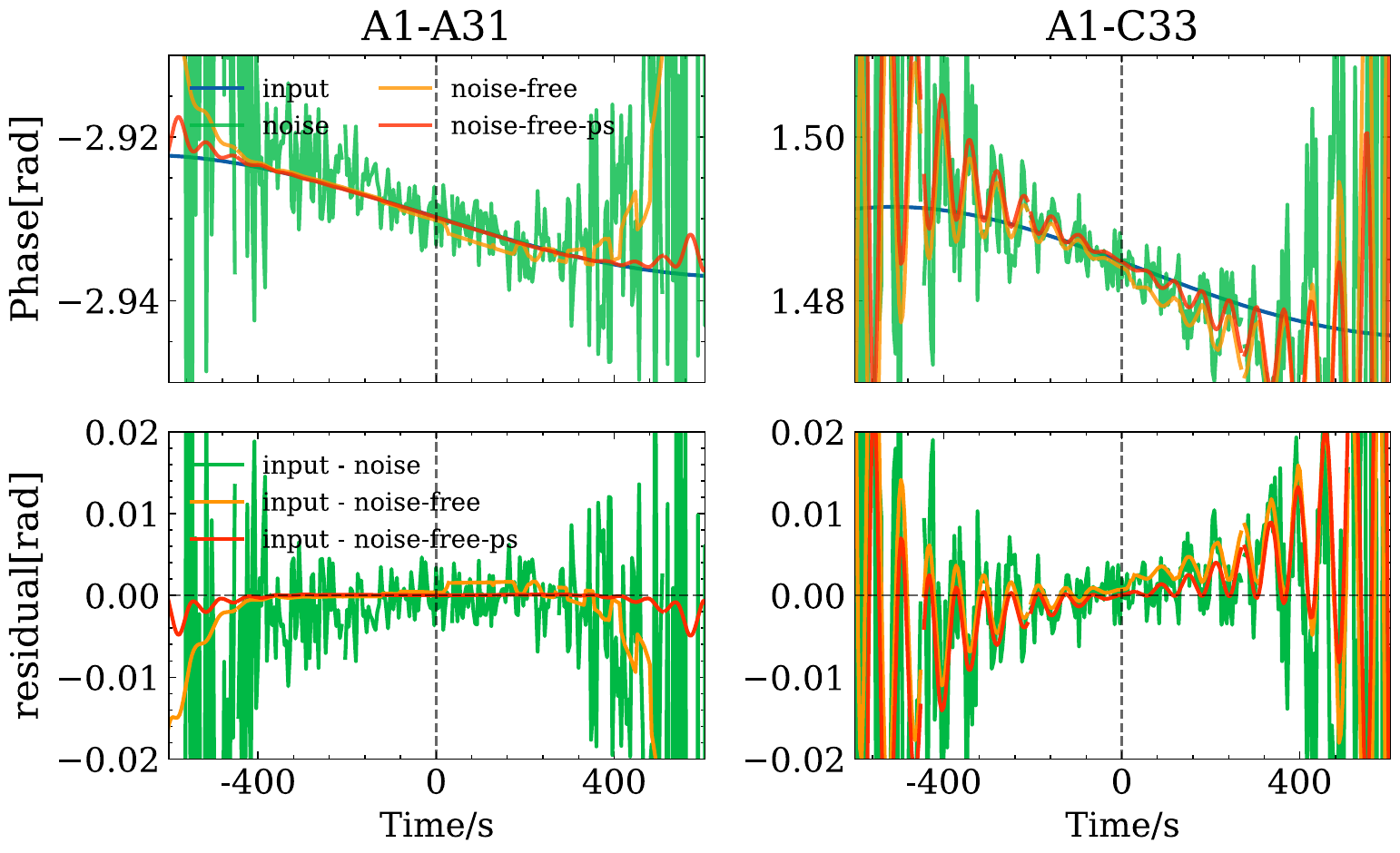}
    \caption{The solution of absolute calibration when transiting Cyg A for the full sky data with/without noise, and the sky with four brightest point sources only of a short baseline (\textit{left} column) and a longer baseline (\textit{right} column). The \textit{top} row shows the input complex gain (blue), the solved gain for the full sky data with noise (green) and noise-free (orange), and for the four point sources sky noise-free data (red). The \textit{bottom} row shows the difference between input and these solved gain.
    }
    \label{fig:ps_cal_full_sky_ps_only}
\end{figure}

After solving $G$ with the eigenvector decomposition method during the transit of bright point sources, we can derive the corresponding complex gains during this period. In our experiment, we have mainly used the Cygnus A (Cyg A) as the calibrator, though Cassiopeia A(Cas A), Virgo A(Vir A) and Taurus A(Tau A) are also used occasionally. It is the brightest source in the sky, and is unresolved by the Tianlai cylinder pathfinder. We plot the input and solved phase for two baselines (A1-A31 and A1-C33) during the Cyg A transit in the top panel of Fig.~\ref{fig:ps_cal_full_sky_ps_only}, and the residue in the bottom panels. In each plot, we show an idealized case where the sky has only a dominating point source and without noise, a case where the sky is given by our fiducial sky model but noise-free, and the more realistic case with noise. 

As we can see from the figure, during the transit of Cyg A, the solutions are basically good within $\pm 300$ seconds of the transit, when the source is within the primary beam and dominates the received signal, though for long baselines we can see some small oscillations at the $10^{-3}$ rad level. Beyond this time scale, the source is outside the primary beam and not dominating, and as a result the solved phase has large errors. Within this window, the model with only point source yields very accurate phase, but for realistic sky model there is still some residue, and with the thermal noise this error is larger.

\begin{figure}
    \centering
    \includegraphics[width=0.8\textwidth]{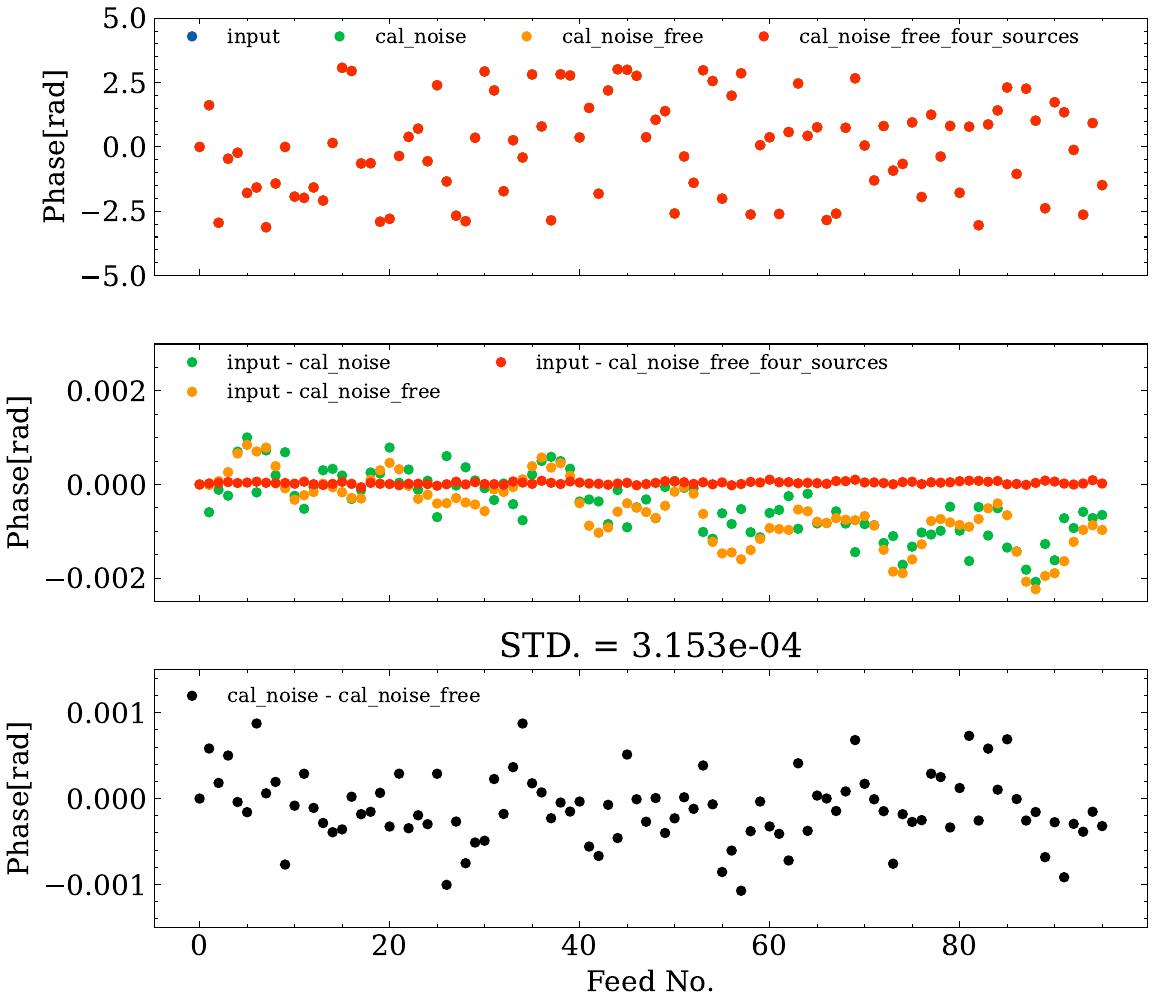}
    \caption{The phase of the complex gains of the feeds as determined by an absolute calibration at the Cyg A transit. \textit{Top}: The phase of the complex gain for the 96 feeds at the time of Cyg A transit. The points show the input phase (blue), the solved phase without noise (orange), the solved phase with noise (green), and the solved noise with simplified sky model of only four strong point sources.
    \textit{Middle}: The residual between the input phase and the solved phase. \textit{Bottom}: The difference between the solved phase for the noise data and the noise-free data. }
    \label{fig:ps_cal}
\end{figure}

The phase and residue for all 96 feeds at the moment of the Cyg A is plotted in Fig.~\ref{fig:ps_cal}.
In the top row of Fig. \ref{fig:ps_cal}, we show the phase of the complex gains for the 96 receivers during the transit, which are randomly distributed in the $(-\pi,\pi)$ range. We show the input, the calibration result without adding noise, and the calibration result with the addition of noise. The differences are small, so in the top panel of Fig. \ref{fig:ps_cal} the points of these cases coincide. The difference between the input and the two calibration solutions are shown in the middle panel of Fig. \ref{fig:ps_cal}. 

As we can see, even in the noise-free case, the calibration solution is still a little different from the input, at the level of $10^{-3} \sim 10^{-4} \rad$. 
To identify the source of this error, we also simulated the case where the sky consists only four well separated points sources (Cyg A, Cas A, Vir A, Tau A), and the visibility is only dominated by one of them at a time. In this simplified case, the solved phase is almost equal to the input. However, once we take the realistic sky radiation model into account, we see there is error in the solved phase, because this contribution is neglected in Equation ~(\ref{eq:ps_vis}) and Equation ~(\ref{eq:eigen_cal}).  We also tested adding the expected thermal noise, this will cause some additional error, but is smaller than the error from over-simplified sky model. In the bottom panel of Fig. \ref{fig:ps_cal}, the difference between the case without thermal noise and with thermal noise is shown, which is at the $10^{-4}$ level. 
For comparison, the closure phase test indicates an error level of $10^{-3}\rad$ for the actual Tianlai cylinder array data\citep{2020SCPMA..6329862L}, comparable or slightly larger than the simulation result. 

Note that for an interferometer array, the size of errors in the calibrated instrument phases depends both on the calibration precision for individual baseline and on the array scale $N$. During the calibration process, each baseline provides an equation to constrain the phase, and the number of baselines scales as $O(N^2)$, while the number of unknown variables scales as $O(N)$. For comparable calibration precision of an individual baseline, the resulting error of the instrument phase achieved with a large array can be much smaller than that with a small array. However, for a very large array this scaling may be broken due to effects not considered in the interferometer model, such as cross-coupling between array elements, or direction-dependent instrument phase, etc., which would limit the precision of the calibration in large arrays. In our present case, the Tianlai cylinder array has an $N \sim 10^2$ which is larger than many previous arrays, so the error is correspondingly smaller. Still, the estimated error size is comparable with what we find in the actual data as indicated by the closure phase measurement, so we believe our present modeling is adequate and captures the main source of error.

\subsection{Relative Calibration}
As the point source calibration could only be made when a known bright point source is transiting, and there are only a few such sources in the sky, we use the relative phase calibration for the rest of the time. 

An artificial CNS is placed nearby to broadcast broadband noise-like signal  periodically. In this simulation, 
we model its visibility during the broadcast as  
\begin{align}
    \nonumber
    V_{ij}^{\text{CNS}} &= S_{\text{CNS}} \frac{A_i(\hat{\vb*{n}}_i) A_j(\hat{\vb*{n}}_j)}{\sqrt{\Omega_i \Omega_j}} \frac{r^2}{r_i r_j} e^{ik(r_i - r_j)} \\ 
    &= C e^{ik(r_i - r_j)},
    \label{eq:CNS_vis}
\end{align}
where $k=2\pi / \lambda$, $r_i$ and $r_j$ are the distance from CNS to feed $i$ and feed $j$. For the real case, this may not be completely accurate, due to near field effect and the reflections from nearby terrain. However, in the relative calibration we do not make any usage of this knowledge, so it should not affect the result.

In the \texttt{tlpipe}, the visibility when CNS is on and its adjacent visibility when CNS is off are assumed to be 
\begin{align}
    V_{ij}^{\text{on}} &= g_i g_j^{*} (V_{ij}^{\text{sky}} + V_{ij}^{\text{CNS}}) + n_{ij}^{\text{on}} \\
    V_{ij}^{\text{off}} &= g_i g_j^{*} V_{ij}^{\text{sky}} + n_{ij}^{\text{off}}
\end{align}
we can assume $n_{ij}^{\text{on}}-n_{ij}^{\text{off}} \approx 0$ when compared with the CNS term, then
\begin{equation}
V_{ij}^{\text{on}} - V_{ij}^{\text{off}} = g_i g_j^{*} V_{ij}^{\text{CNS}} 
     \label{eq:CNS_vis2}
\end{equation}
As the CNS and feeds are fixed, the phase change in its visibility would come from the variation of the complex gain. Regardless the exact CNS visibility, as long as it does not change, 
the phase of the observed visibility is 
\begin{align}
\Phi_{ij}=\text{Arg}(V_{ij}^{\text{on}} - V_{ij}^{\text{off}}) \approx \text{Arg}(g_i g_j^*) + \text{Arg}(V_{ij}^{\text{CNS}})
    \label{eq:rel_cal1}
\end{align}
and its change depends only on the variation of the complex gain and does not depend on the exact form of the CNS induced visibility, 
\begin{equation}
\Delta \Phi_{ij}(t) \equiv \Phi_{ij}(t)-\Phi_{ij}(t_0)
\label{eq:rel_cal2}
\end{equation}
where $\Phi_{ij}(t)$ is the phase at time $t$, and 
$t_0$ is the time of the last absolute calibration.
We can then calibrate the visibility as 
\begin{equation}
    V_{ij}^{\text{rel-cal}} = e^{-i\Delta \Phi_{ij}}V_{ij}^{\text{off}} 
    \label{eq:rel_cal3}
\end{equation}

\begin{figure}[ht]
\centering
\includegraphics[width=0.9\textwidth]{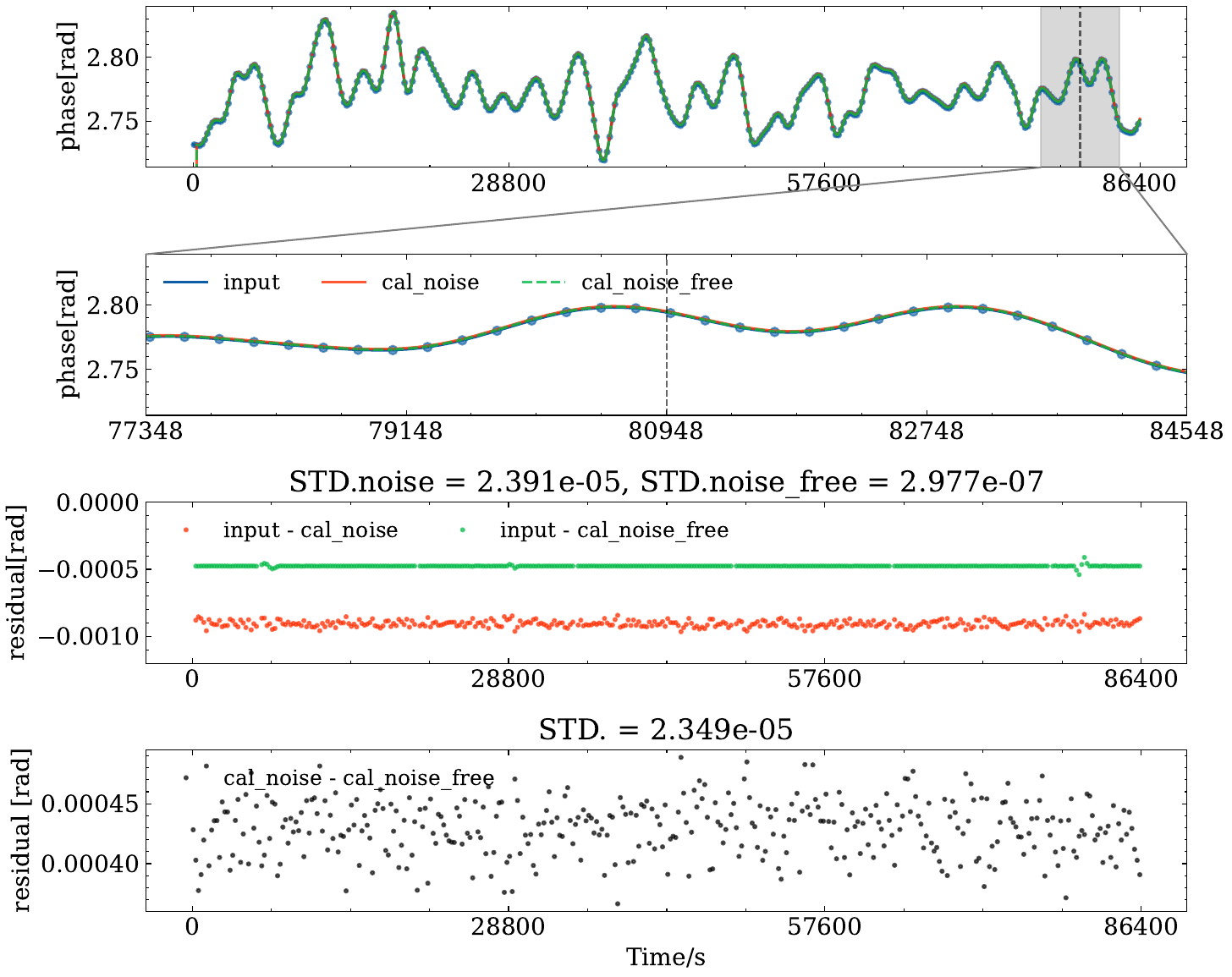}
\caption{The solution to the phase variation and the residual after calibration for one baseline(A1-A21 in this case). The dots in all panels indicate the on-time of CNS. \textit{Top Row }: the solved phase for noise data(orange) and noise-free data(green dashed)for a whole day data compared with the input(blue). \textit{Second Row}: Zoom in on the period when Cyg A is transiting. \textit{Third Row}: The residual between the input and the calibration solution. \textit{Bottom Row}: The difference between the solved phase of the two case, the variation of which is from the noise. 
}
    \label{fig:gain_result_1}
\end{figure}

We can estimate the minimum magnitude of the error of relative calibration as induced by the measurement noise: for the complex visibility of our unpolarized case, the measurement error in each of the real and imaginary component is about $\sigma_{\rm noise} / \sqrt{2} \approx 0.0022 \K \cdot \sr$.
In the absolute or relative calibration, if the visibility induced by the calibration source is 
$V$, and $|V| \gg \sigma$, then the measured visibility has a phase error of 
\begin{equation}
\delta \phi \approx \frac{\sigma}{|V|} 
\end{equation}

The CNS signal strength is about $10^2 \K \cdot \sr$ in our simulated visibility 
so the error $\delta \phi_{\rm rel}$ induced by the thermal noise of measurement is 
\begin{equation}
  \delta \phi_{\rm rel} \sim \frac{0.0022}{100} \approx 2.2 \times 10^{-5} \rad
\end{equation}

Fig. \ref{fig:gain_result_1} shows the phase of the complex gain of a typical baseline after the two-step (absolute+relative) calibration. In the top panel we show the phase evolution over one day, and in the second row show the phase during the transit of Cyg A. Both the input and calibration results (with noise or without noise) are shown. In the third row, we show the difference between the input and the calibration solution. Here we plot two cases: the case with noise in both the absolute and relative calibration; and the noise-free case, where we ignore the noise in both the absolute and relative calibration.
As noted before, the solution has a phase error due to numerical error at a level of $10^{-4} $rad, which is set largely by the absolute calibration. Due to this error from the absolute calibration, the two solutions are different from the beginning, but this difference is almost constant, as the phase variation is precisely corrected by the relative calibration procedure. 
Finally, in the bottom row, we show the difference between the solution with and without noise, this shows an error of about $2.3 \times 10^{-5}$, which is consistent with our estimates of relative calibration error from above.  
In addition to the error induced by thermal noise, there could be other errors, e.g. from reflection and multi-path effects, which we will ignore in the present work.  

Our simulation result shows that the absolute calibration could fix the randomly distributed instrument phase, and for the daily instrumental phase variations at the $10^{-2}\rad$ level, the relative calibration could reduce it significantly. However, it also shows that during the absolute calibration, there is an error at the $10^{-4} \sim 10^{-3} \rad$ level, which will persist subsequently. This error is partly due to the fact that the sky is actually not dominated by a single radio source, and partly due to thermal noise. The absolute calibration accuracy might be improved by jointly fitting the phase with the data obtained during the whole transit process instead of a single epoch data at the transit point, and also perhaps by adopting more sophisticated sky model.

The relative calibration would track how the phase vary from the initial phase, but would not improve on the initial phase determination. The error in the relative calibration is however smaller, at the $10^{-5}$ level.

\section{Map-making Results}
\label{sec:map-making}
The output of an interferometer array is the visibility data $V_{ij}$, from this data the sky brightness temperature map $T(\hat{\vb*{n}})$ can be obtained by solving the inverse problem of the visibility Equation \eqref{eq:vis_eq1}\citep{2017MNRAS.465.2901Z}.
While map-making is not absolutely necessary for obtaining the 21 cm power spectrum, which can in principle be obtained directly from the visibility data, a map is more intuitive than the visibilities, so it provides a good check for potential problems and systematic effects in the observation.

For drift scan surveys, where the telescope is fixed on the ground and sky drift by due to the rotation of Earth, an $m$-mode decomposition method can be used, which simplifies the problem and reduces the amount of computation significantly
\citep{2014ApJ...781...57S, 2015PhRvD..91h3514S,2016MNRAS.461.1950Z,2016RAA....16..158Z}. In the present work we shall simulate the map-making processing using the $m$-mode formalism. 

\subsection{The $m$-mode formalism}
\label{sec:mm_mmode}

The $m$-mode formalism works with the drift scan telescope, and has been used in the sky reconstruction for the Tianlai dish and cylinder arrays using simulated data\citep{2016MNRAS.461.1950Z,2016RAA....16..158Z}, and the analysis of OVRO-LWA data\citep{2018AJ....156...32E, 2019AJ....158...84E}. We summarize it briefly below.

Take the fact that the measured visibility changes periodically with the sidereal day as the Earth rotates, we can replace the time dependence of visibility with the dependence on the sidereal hour angle $\varphi$ as $\varphi(t)$, then $V_{ij}(t)$ and $B_{ij}(\hat{\vb*{n}}; t)$ in Equation \eqref{eq:vis_eq1} and \eqref{eq:btf} is expressed as $V_{ij}(\varphi)$ and $B_{ij}(\hat{\vb*{n}}; \varphi)$. Taking advantage of the periodicity of $\varphi$, we can Fourier transform $V_{ij}(\varphi)$ with respect to the $\varphi$ as
\begin{equation}
    V_{m}^{ij} = \int^{2\pi}_{0} \frac{\dd \varphi}{2\pi} V_{ij}(\varphi) e^{-im\varphi} \\
    \label{eq:mmode1}
\end{equation}
where $V_{m}^{ij}$ is the $m$-mode indexed $m$, the Fourier conjugate variables to the sidereal hour angle $\varphi$. The $m$-modes $V_{m}^{ij}$ correspond to the components of the visibility $V_{ij}(\varphi) $ varying on different timescales, larger values of $m$ correspond to the faster varying components. $V_{0}^{ij}$ is the average of the visibility over sidereal time.

Take the spherical harmonics transform of the beam transform matrix $B_{ij}(\hat{\vb*{n}}; \varphi)$ and the sky brightness temperature $T(\hat{\vb*{n}})$,
\begin{align}
    B_{ij}(\hat{\vb*{n}}; \varphi) &= \sum_{lm} B_{lm}^{ij}(\varphi) Y_{lm}^{*}(\hat{\vb*{n}})
    \label{eq:btf_m} \\
    T(\hat{\vb*{n}}) &= \sum_{lm}a_{lm}Y_{lm}(\hat{\vb*{n}})
    \label{eq:sky_m},
\end{align}
substitute them into Equation \eqref{eq:vis_eq1} and combine with Equation \eqref{eq:mmode1}. Ignore  the complex gain for now, we have
\begin{equation}
    V_{m}^{ij} = \sum_{lm'}\int \frac{d\varphi}{2\pi} B_{lm'}^{ij}(\varphi) a_{lm'} e^{-im\varphi} + n_{m}^{ij}.
    \label{eq:mmode2}
\end{equation}
Using the property of the drift scan telescope, of which the pointing changes with the rotation of Earth, the beam transfer function can be written as $B_{lm'}^{ij}(\varphi) = B_{lm'}^{ij}(0)e^{im'\varphi}$. Substitute it into Equation \eqref{eq:mmode2}, it gives
\begin{equation}
    V_{m}^{ij} = \sum_l B^{ij}_{lm} a_{lm} + n_{m}^{ij}.
    \label{eq:mmode3}
\end{equation}

For any $m > 0$, the positive and negative $m$-modes $V_{m}^{ij}$ and $V_{-m}^{ij}$ are independent data. For the real-valued sky temperature, $a_{l, -m} = (-1)^{m} a_{lm}^{*}$, then 
\begin{equation}
    V_{-m}^{ij*} = \sum_l (-1)^{m}B^{ij*}_{l,-m} a_{lm} + n_{-m}^{ij*}.
\end{equation}
Equation \eqref{eq:mmode2} can be rewritten in matrix form
\begin{equation}
    \vb*{v = Ba + n},
    \label{eq:mmode5}
\end{equation}
where $\vb*{v, B, a, n}$ are the matrix notation for the  $V_m^{ij, \pm}, B^{ij, \pm}_{lm}, a_{lm},  n_{m}^{ij, \pm}$ respectively. It is noted that $\vb*{B}$ is a block diagonal matrix, which allows the 
operation on Equation \eqref{eq:mmode5} can be performed $m$-by-$m$, so the operation on large matrices can be split into smaller ones.

Due to the fixed size of the telescope, its sensitivity to larger $l$ and $m$ decreases rapidly beyond a scale, which reduces Equation \eqref{eq:mmode3} to a finite sum. We limit the largest measurable spherical harmonic degree $l$ by the largest dimension of the array $D_{\text{max}}$, and $m$ is determined by the largest distance in the E-W direction $D_{\text{E-W}}$, which are taken as $l \lesssim 2\pi D_{\text{max}} / \lambda$ and $m \lesssim 2\pi D_{\text{E-W}} / \lambda$. For Tianlai cylinder pathfinder at 750 MHz, we take $l_{\text{max}} = 742$ and $m_{\text{max}} = 715$ in our simulation.

The least-squares solution to Equation \eqref{eq:mmode5} which minimize $\norm{\vb*{v - Ba}}^2$ is
\begin{equation}
    \hat{\vb*{a}}_{\text{LS}} = (\vb*{B^{*}B})^{-1}\vb*{B^{*}v},
    \label{eq:ls_solver}
\end{equation}
as there are unmeasured modes of the sky due to incomplete $uv$-coverage, the rank of matrix $\vb*{B}$ is usually not full, therefore the matrix $(\vb*{B}^{*}\vb*{B})$ is not invertible. A solution of least-squares solution can be written in terms of the Moore-Penrose pseudo-inverse(denoted with $+$) as
\begin{equation}
     \hat{\vb*{a}}_{\text{LS}} = \vb*{B}^{+} \vb*{v}.
    \label{eq:ls_solver2}   
\end{equation}
The Moore-Penrose pseudo-inverse regularizes these singular modes (singular values below a threshold) to zero. The threshold is chosen to ensure a good reconstruction. For maps with low noise, low threshold could be adopted for the inversion. It would be determined by some experiment.

\subsection{Map Reconstruction}
\label{sec:map_recon}
\begin{figure}[htb]
    \centering
    \includegraphics[width=0.48\textwidth]{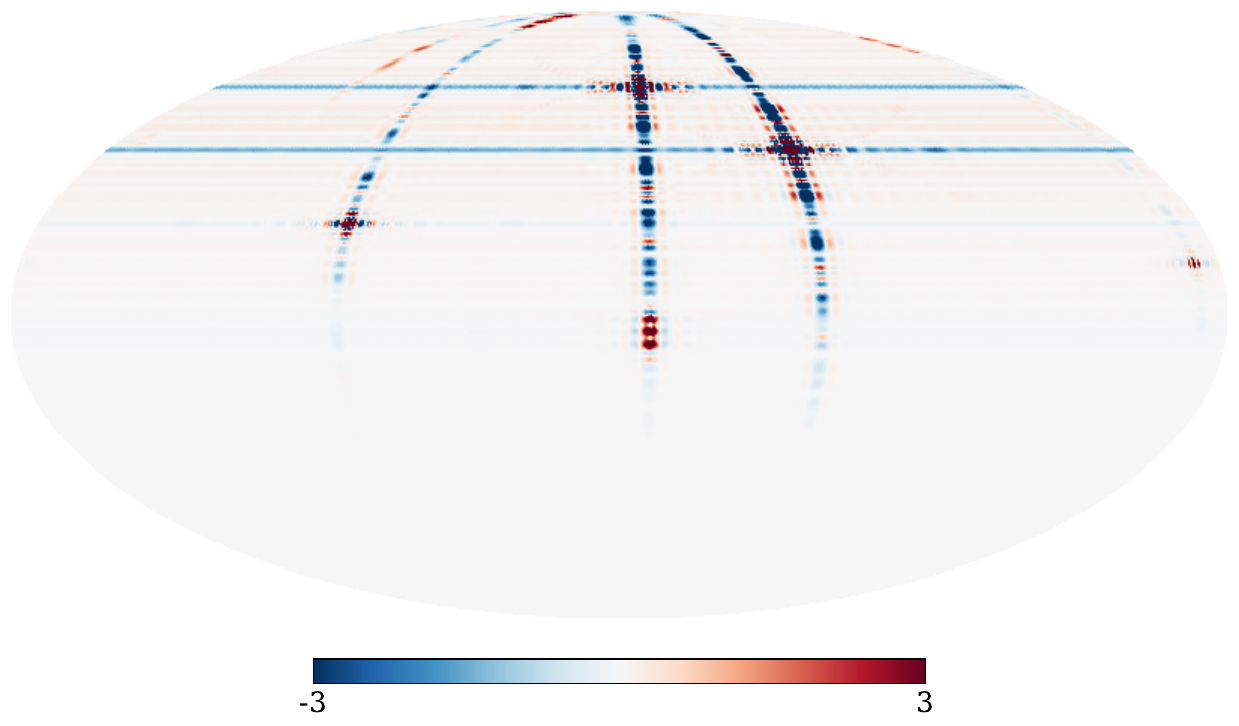}
    \includegraphics[width=0.48\textwidth]{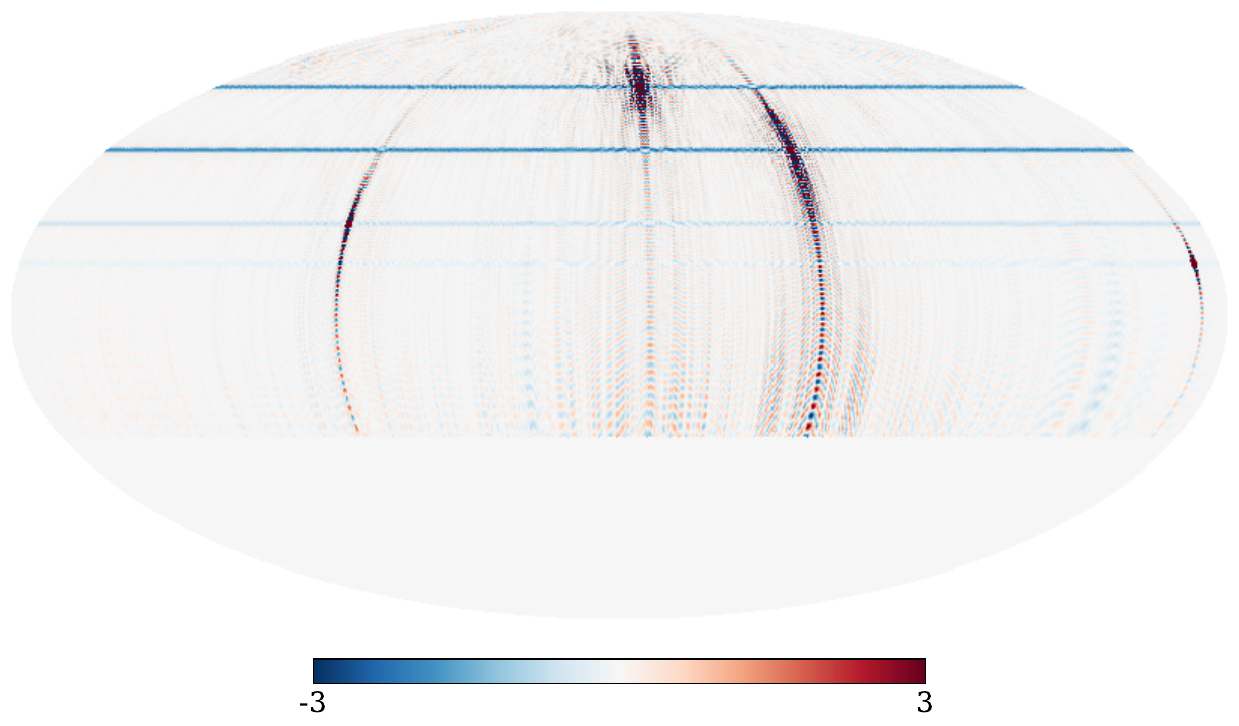}
    \caption{The dirty map (left) and reconstructed map (right) for the case of four bright sources (except the bright sources, all sky radiation is set to zero in the input map). The maps have been smoothed to an approximate resolution of Tianlai cylinder pathfinder in the EW direction with a Gaussian beam filter of 0.0133 rad. }
    \label{fig:map_four_sources}
\end{figure}

To test this algorithm, we first generate a simple sky map, in which all pixels are set to zero, except for a few bright points placed at different places in the sky, and run the simulation. In the left panel of Fig. \ref{fig:map_four_sources}, we show the dirty map, which is obtained by \red{A typo in the equation}
\begin{equation}
\vb*{a}_{\rm dirty} = \vb*{B}^{*} \vb*{v}.
\end{equation}
In the right panel of Fig. \ref{fig:map_four_sources}, we show the reconstructed map obtained from Equation~(\ref{eq:ls_solver2}), which is smoothed to an approximate resolution of Tianlai cylinder pathfinder in the EW direction with a Gaussian beam filter with ${\rm FWHM}\sim 0.3997 \m/30\m \approx 0.0133 \rad$, corresponding to the maximum resolution of the beam in the East-West direction.  

The dirty map reflects the response of the array for the drift scan observation. We can see that there are very extended sidelobes along the North-South direction, and also shorter sidelobes in the East-West direction at each of the bright radio sources. These are basically along the longitude and latitude lines, due to the structure of the cylinder array and its drift scan mode of observation. The sidelobes have complicated structures, which would be frequency-dependent, and can affect the extraction of the 21cm signal. 

In the reconstructed map, the instrument response is deconvoluted, though it is not perfect and depends on the deconvolution algorithm. We see that in the reconstructed image, the sidelobes are not as strong as in the dirty map, especially the sidelobes along the East-West direction are significantly reduced. However, the sidelobe is not completely eliminated, especially in the North-South direction. Compared with the dirty map, the position of the sidelobe actually shifted, showing that while reducing the sidelobe in some parts of the map, the imperfect deconvolution may also introduce new fake features at other parts of the map.  

There are also lines along the whole latitude lines passing through the bright sources, but they are not sidelobes. These are produced because we have subtracted the $m=0$ modes, which do not vary and would be subtracted as correlated noise in the real data. However, this means practically that the average along each latitude must be zero, so for the latitudes where the bright sources are located, the other places are set to a negative value.  

\begin{figure}[t!]
    \centering
    \includegraphics[width=0.55\textwidth]{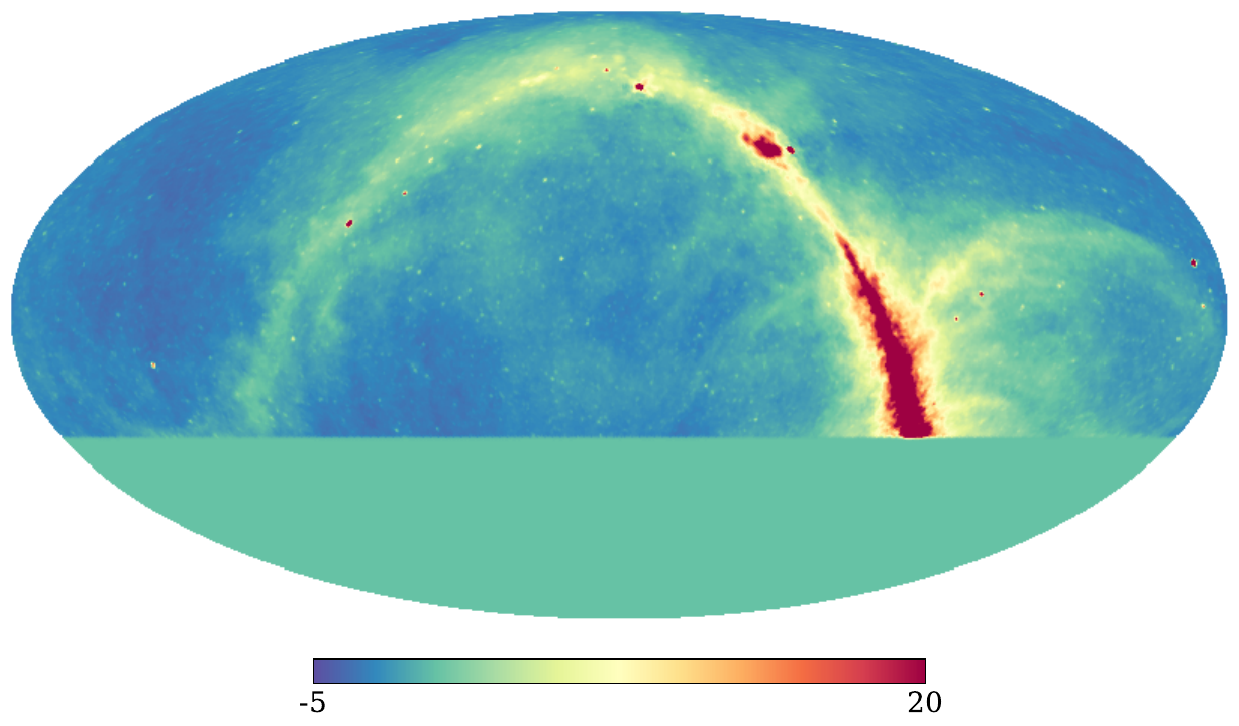}
    \includegraphics[width=0.48\textwidth]{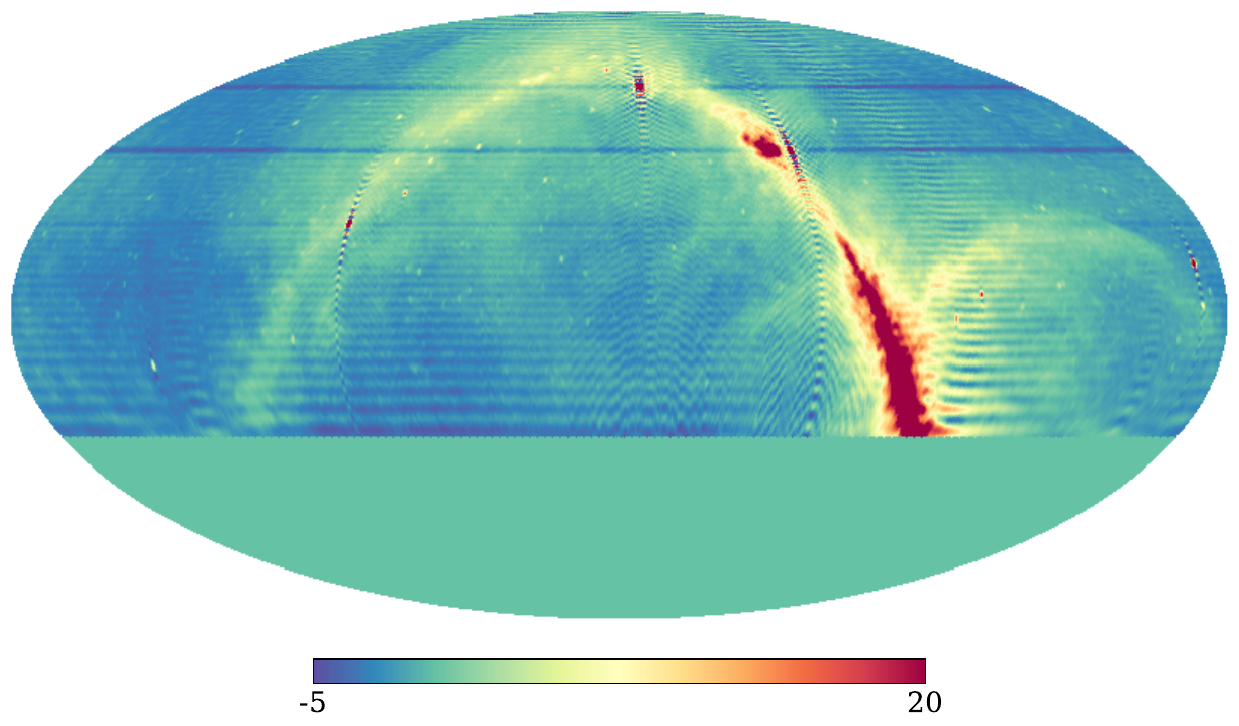}
    \includegraphics[width=0.48\textwidth]{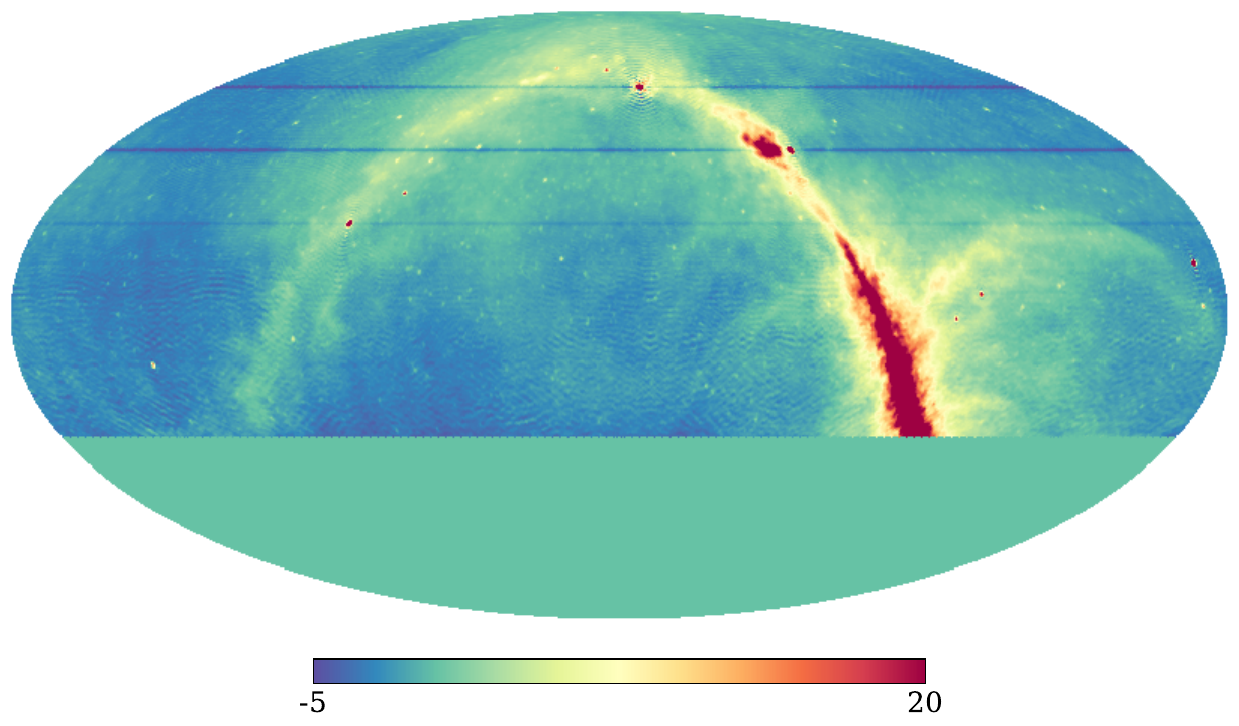}
    \caption{The filtered input sky map(\textit{top}) and the reconstructed sky maps(\textit{bottom}) in Molleweide project, the area below south of $-30^{\circ}$ which is close to or below the horizon of the location of Tianlai array is masked.
    \textit{Bottom left}:  the reconstructed map from the data without gain variation and noise with threshold $\epsilon = 10^{-3}$; \textit{Bottom right}: $\epsilon = 10^{-5}$. All maps have been smoothed as we do in Fig. \ref{fig:map_four_sources}.
   }
    \label{fig:reconstructed_map}
\end{figure}
We then generate an input map from the sky model, and also reconstructed maps with the mock data.  In this work, we set the Moore-Penrose pseudo-inverse threshold to be $\epsilon \times \text{max}(\sigma_i)$, where $\sigma_i$ is the singular values of matrix $\vb*{B}$, and $\epsilon$ is a preset threshold value, which we take here as $10^{-3}$. The resulting maps are plotted in Fig.~\ref{fig:reconstructed_map} in the Molleweide projection,
the area below south of $-30^{\circ}$ is mask in consideration of the field of view of Tianlai Array.

As our telescope is sensitive only to modes with $l<600$ (this is a conservative number, it is likely smaller than this), even with perfect reconstruction we will still loose the $l>600$ modes, we therefore compare with a map made from the original sky map shown in Figure \ref{fig:skymap}, but with all modes of $l>600 $ removed, which we shall refer to as the (low-pass) filtered input map, even though the actual input map which we use to generate the visibilities does have the $l>600 $ modes. This is done by first computing the spherical harmonics coefficients of it, setting those components with $l > 600$ to zero, and transform back. Similarly, we also discard all the $m = 0$ modes, which are prune to error as they do not vary over time. This also removes the $l=0$ mode, i.e. the global mean temperature of the map. As a result, some pixels of the map become negative. This map is shown in the top panel of Fig.~\ref{fig:reconstructed_map}.

In the bottom left of Fig.~\ref{fig:reconstructed_map} we show the reconstructed sky maps in the ideal case with threshold $\epsilon = 10^{-3}$, i.e. from the mock visibilities generated by the sky model, without any noise or gain variation. The reconstructed map is also smoothed with a Gaussian filter with ${\rm FWHM} \sim 0.0133 \rad$.
As we can see, there are differences from the input map. Besides the longitudinal sidelobes and negative latitude lines around the bright points which we noted in Fig.~\ref{fig:map_four_sources}, another notable feature is the horizontal stripes extended from the brighter part of the galactic plane, which form a comb-like structure. 

\begin{figure}[htb]
    \centering
    \includegraphics[width=0.48\textwidth]{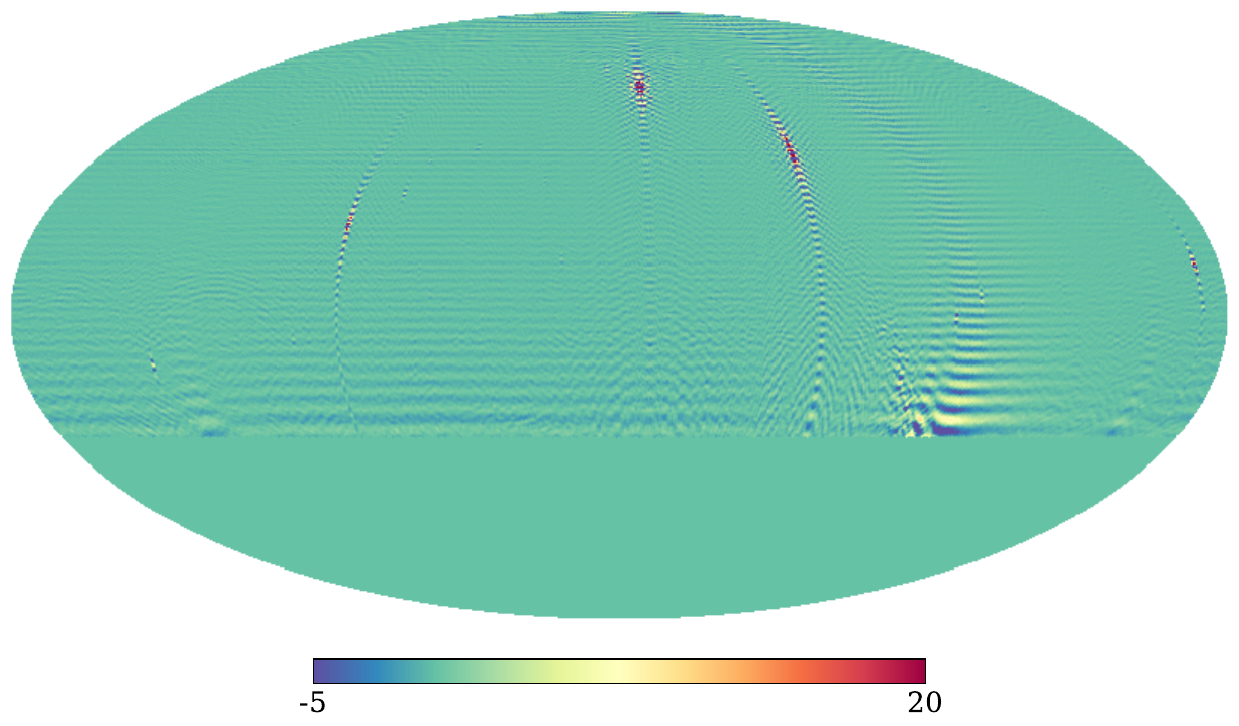}
    \includegraphics[width=0.48\textwidth]{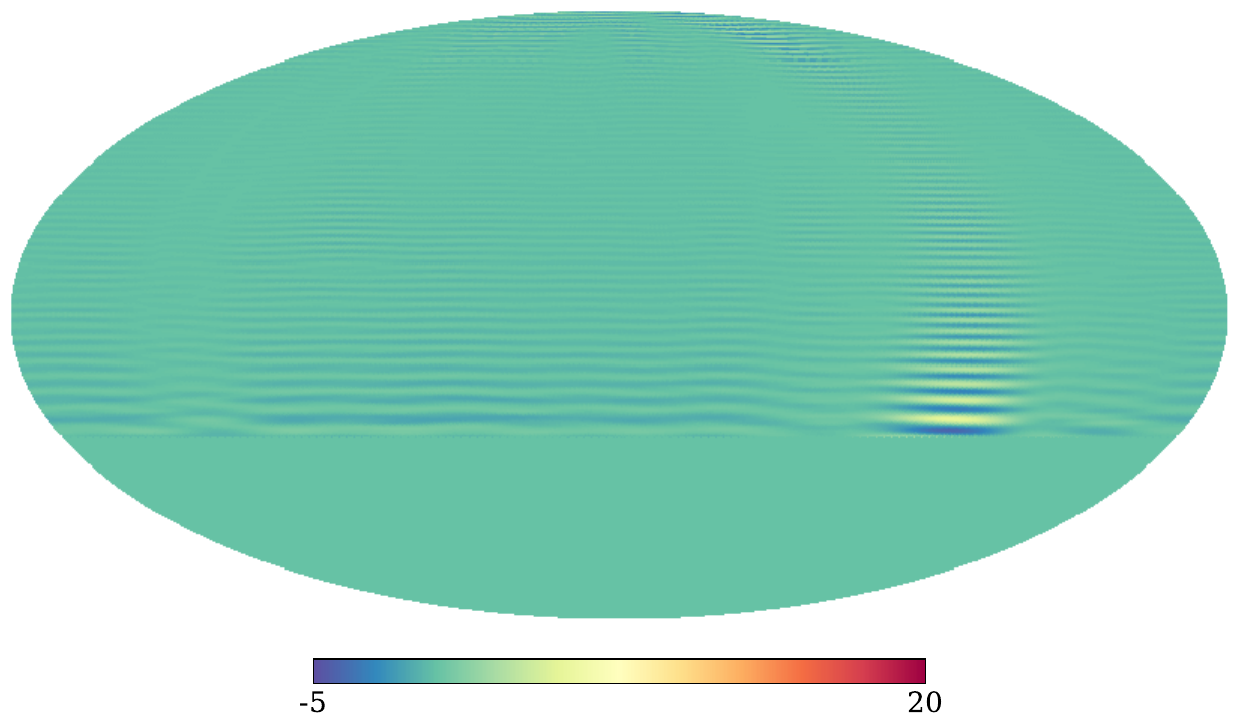}
    \caption{\textit{Left}: The difference of the two maps in the bottom of Fig. \ref{fig:reconstructed_map}; \textit{Right}: The difference of the two reconstructed maps with the $0 < m < 10$ and $l < 200$ modes only. 
   }
    \label{fig:res}
\end{figure}

In the bottom right of Fig.  \ref{fig:reconstructed_map}, we plot the reconstructed map in the ideal case with a smaller threshold $\epsilon=10^{-5}$, for this case, the comb-like artifact near the galactic plane disappears, and generally a more accurate map is reconstructed.
This is because when we impose the cut off threshold, modes with small eigenvalues or singular values are discarded. These modes however also contain some information about the sky. Comparing with the input map, setting these modes to zero is equivalent to adding their negative. Unlike modes generated by noise, however, these modes are regular and show up clearly in the map. In the left panel of Fig. \ref{fig:res}, we plot the difference between the maps of the two thresholds. The differences show most clearly near the brightest part of the sky, i.e. near the bright point sources, and the brightest part of the galactic plane.
The comb-like structure are mainly due to missing small $m$ (e.g. $m < 10$) modes caused by the pseudo-inverse truncation. If we denote these missing small $m$ (e.g. $m < 10$) modes as $\Delta a_{lm}$, the missing map component would be $\sum_{lm} \Delta a_{lm} Y_{lm}(n)$. We  plot the map made with only the $m < 10$ and $l < 200$ modes in the right panel of Fig. ~\ref{fig:res}, and one can see clearly the comb-like structure. On the other hand, the structure around the few bright point sources do not show up in this case, which shows that they are generated mostly by the high $m$ modes. 

\begin{figure}
  \centering
  \includegraphics[width=0.6\textwidth]{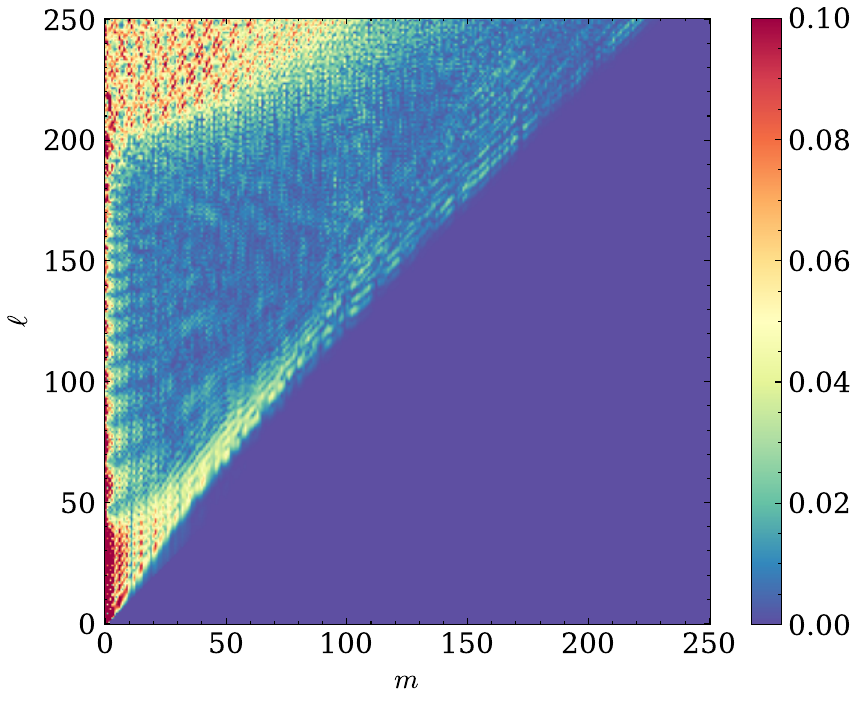}
  \caption{The amplitude of the  spherical harmonics $a_{lm}$ of the difference map.}
  \label{fig:res_alm}
\end{figure}

We also show the spherical harmonic coefficients $a_{lm}$ for the difference map of $\epsilon=10^{-3}$ and $\epsilon=10^{-5}$ case
in Fig. \ref{fig:res_alm}. As can be seen from the figure, the largest $a_{lm}$ of this differential plot are those $m < 10$ modes, concentrated near the $\ell$-axis. We expect such small $m$ modes will produce sizable features along the latitude lines, which would appear as the comb-like structure we see in the reconstructed map with $\epsilon=10^{-3}$.  

From the above analysis, we see that a smaller $\epsilon$ is desirable in the reconstruction.  
However, in the presence of noise, the value of $\epsilon$ can not be too small, otherwise the reconstructed map will be affected greatly by the noise, often rendering the whole solution unusable. The threshold $\epsilon=10^{-3}$ ensures that even with relatively large noise, for example those for a single day observation, the reconstruction could still proceed smoothly. However, if the noise in the visibility could be reduced by adding the the data collected at the same local sidereal time (LST) on different days, it may be possible to impose smaller $\epsilon$ value and obtain more accurate maps. 

\subsection{Calibration Error and Noise}
\begin{figure}[ht]
    \centering
    \includegraphics[width=\textwidth]{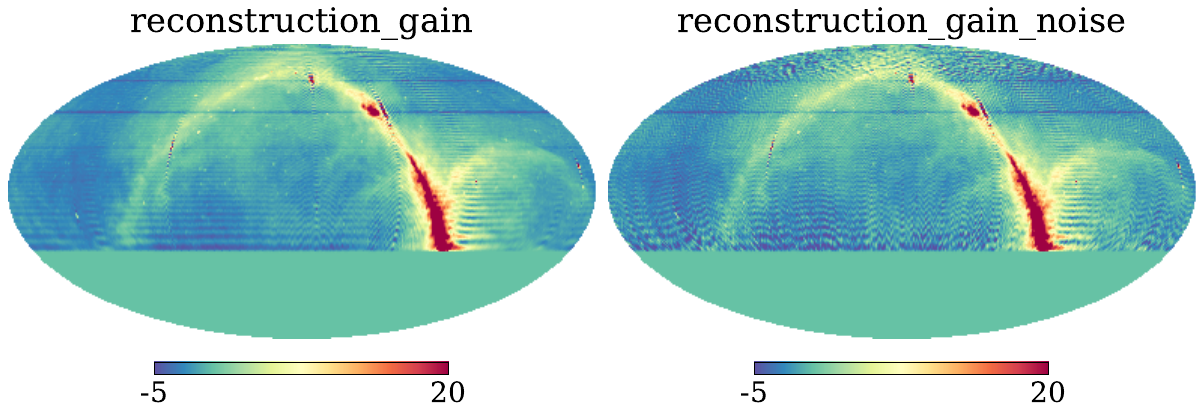}
    \caption{ \textit{Left}: the reconstructed map with complex gain variation after calibration. 			\textit{Right}: the reconstructed map with complex gain variation and noise after calibration. 
   }
    \label{fig:reconstructed_map_gain+noise}
\end{figure}

With the map-making error understood, we now consider the impact of gain variation and noise. We  use the method outlined in Sec.\ref{sec:tlpipe} to calibrate the gains, and use the calibrated visibilities to make the reconstructed map, as shown by the left panel of Fig.~\ref{fig:reconstructed_map_gain+noise}. This map is very similar to the reconstructed map in the ideal case shown in Fig.~\ref{fig:reconstructed_map}. We also add noise, and run through the same procedure in the presence of noise, and the result is shown in the right panel of Fig.~\ref{fig:reconstructed_map_gain+noise}. The noise in the visibility measurement also translates to fluctuations in the map.

\begin{figure}[h]
    \centering
    \includegraphics[width=\textwidth]{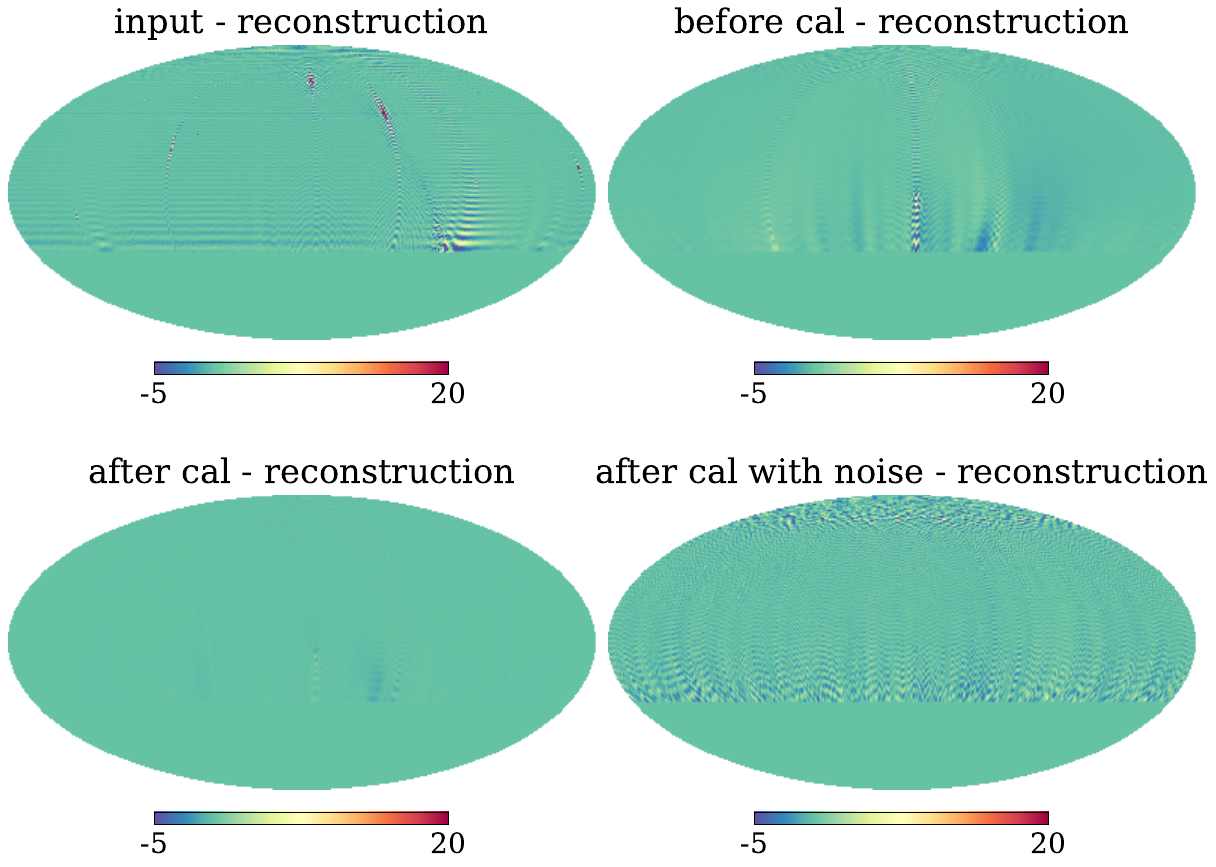}
    \caption{The difference between reconstructed maps. }
    \label{fig:reconstructed_map_diff}
\end{figure}

To see the impact of these more clearly, we plot the differences of these maps in Fig. \ref{fig:reconstructed_map_diff}, using the same color scale. In the top left we show the difference between the reconstructed map in the ideal case and the input map. As noted in Sec.\ref{sec:map_recon}, there are large differences near the bright sources, and we can also see the horizontal stripes including the comb-like structure near the galactic plane. 

In the top right panel, we plot the difference between the reconstruction with gain variation but without making the relative calibration. The differences are small, as the gain variations are not large by itself. Indeed, we found from the actual observational data during night that even without the relative calibration, we can still make pretty good maps. In the current simulation we have adopted a gain fluctuation model which have fluctuation level similar to the real case, so it is not surprising to find this result. The residue difference in the map are again located mainly at the bright spots, or their sidelobes which are along the longitude and latitude lines. Once relative calibration is done,  these residue differences also disappear (bottom left), showing that the relative calibration can effectively mitigate this error.  

Finally, in the bottom right, we show the difference in the reconstructed map with both calibrated gain variation and thermal noise. The impact of the noise is uniformly distributed across all right ascensions, but as we can see from the figure it is not uniform in the North-South direction. All the full-sky maps are computed using HEALPix and illustrated with Mollweide projection, thus it is not due to unequal pixel area. The noise is lowest around mid-latitude ($\sim 45^\circ$), and increases toward south and north. This suggests that it is due to the sensitivity of the telescope, as it has the best sensitivity for the part of the sky which is right above the telescope, but the sensitivity degrades on the two sides.  Note here we plotted the case for the error of one day's data, and its magnitude could be reduced by stacking more days of observations.

\subsection{Angular Power Spectrum}
\label{sec:angularpower}
Finally, we assess how these errors would affect the angular power spectrum, which is defined as
$C_{l} \delta_{l l'} \delta_{mm'} \equiv \left< a_{l m} a_{l' m'}^*\right>$,
and under the ergodic assumption,  
\begin{equation}
    C_{l} = \frac{1}{2l + 1}\sum^{l}_{m=-l} |a_{l m}|^2.
\end{equation}
An optimal estimation of the angular power spectrum for incomplete sky coverage has been discussed in \citet{2019MNRAS.484.4127A}. Here, we simply calculate the angular power spectrum of the input map and the reconstructed map by masking the part south of $-30^{\circ}$ latitude.

\begin{figure}[htb] 
    \centering
    \includegraphics[width=0.49\textwidth]{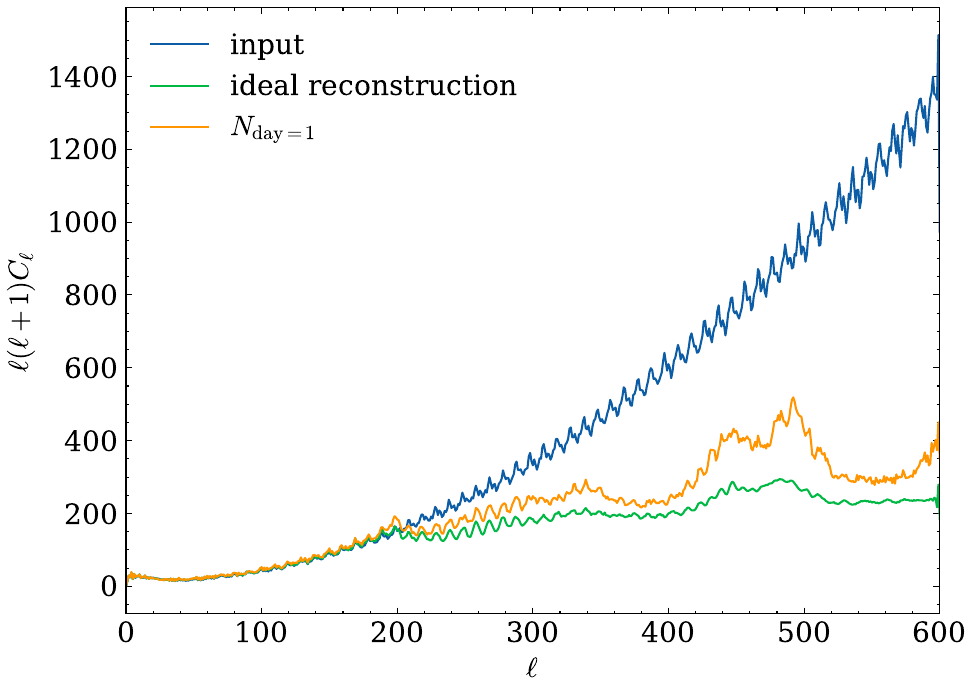}
    \includegraphics[width=0.49\textwidth]{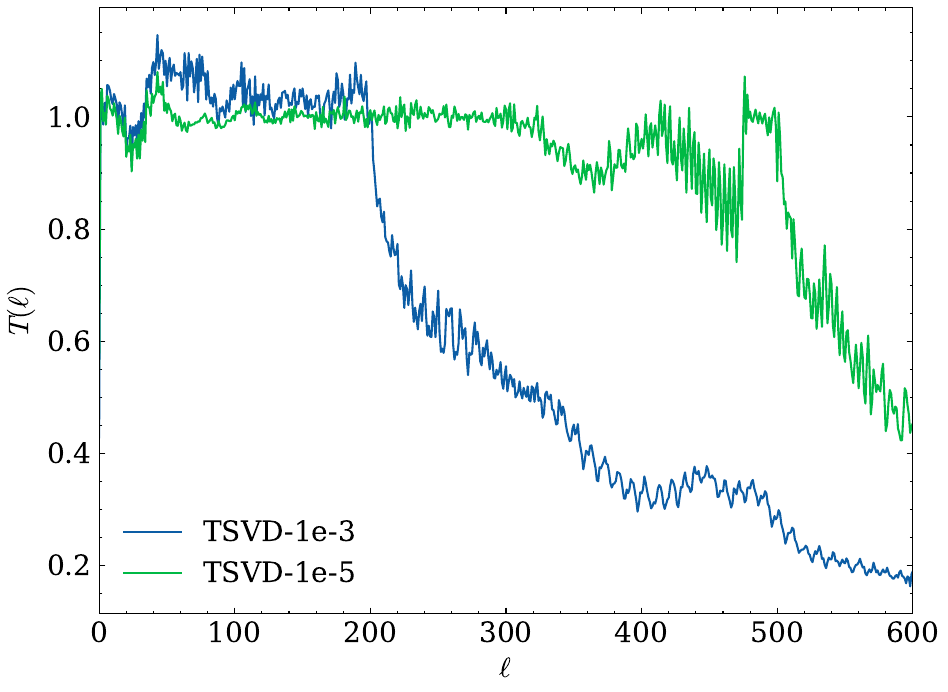}
    \caption{\textit{Left}: The angular power spectrum for the input(blue) and reconstructed maps from the data with complex gain variation and noise(orange), and the ideal case(green, i.e. calibrated perfectly, noise-free) with $\epsilon = 10^{-3}$. \textit{Right}:
    The transfer function  $C_{\ell}/C_{\ell}^{\rm input}$ for the ideal case with $\epsilon=10^{-3}$ and $10^{-5}$.
    }
    \label{fig:angular_ps}
\end{figure}

In the left panel of Fig. \ref{fig:angular_ps} we plot the angular power spectrum for the input map (blue curve), for the reconstructed map in the ideal case (green curve), and the reconstructed map with calibrated gain variation and a noise corresponding to one day observation (orange curve), where $\epsilon = 10^{-3}$ is applied when reconstructing the maps. The wiggles on the angular power spectrum are produced by some point sources in the sky.  In the right panel of Fig.~\ref{fig:angular_ps} we plot the transfer function $T(l)=C_l/C_l^{\rm input}$, which quantifies the amount of loss in the angular power spectrum under ideal condition for $\epsilon=10^{-3}$ and $10^{-5}$.

We can see that generally, at $l \lesssim 200$ the angular power spectrum is well recovered, and the transfer function $T(l) \sim 1$ for both $\epsilon$ values, which means that most of the modes are well recovered, though there are fluctuations due to the reconstruction errors discussed in Sec.\ref{sec:map_recon}. At $l \gtrsim 200$, however, for our fiducial case of $\epsilon=10^{-3}$, the angular power spectrum falls systematically below that of the input map, and the transfer function drops significantly below 1, showing that at such angular scales and precision the spherical harmonics are not well recovered. For the $\epsilon=10^{-5}$ case, the transfer function remains flat at $l \gtrsim 200$, only begin to decline significantly at $l \gtrsim 500$. This is because the discarded modes have small singular values, which would be unrecoverable in the presence of noise. 

Note that the mode probed by a baseline $b$ is $l \lesssim 2\pi b/\lambda$. For our array, the longest baseline in the North-South direction is about 12.4m, corresponding to $l \sim 195$. Thus, the array could not probe some of the modes at $l>195$ in the North-South direction, and in reconstruction such modes will be missing, reducing the total power. This may be the reason why the angular power spectrum of the reconstructed map falls systematically below that of the input.

\begin{figure}[ht!]
    \centering
    \includegraphics[width=0.49\textwidth]{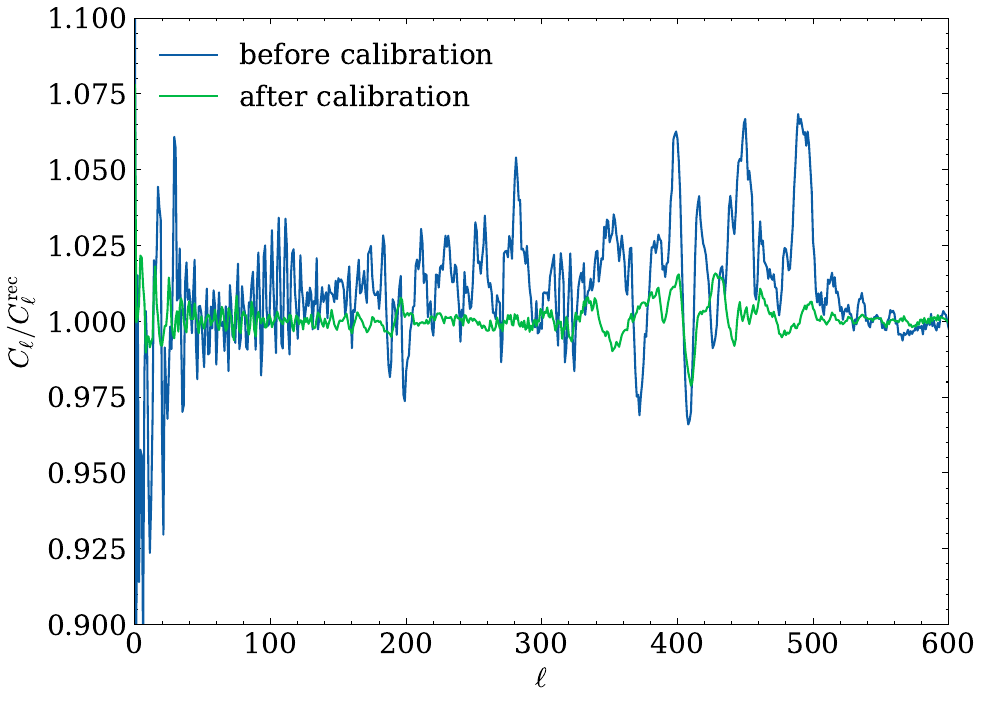}
    \includegraphics[width=0.49\textwidth]{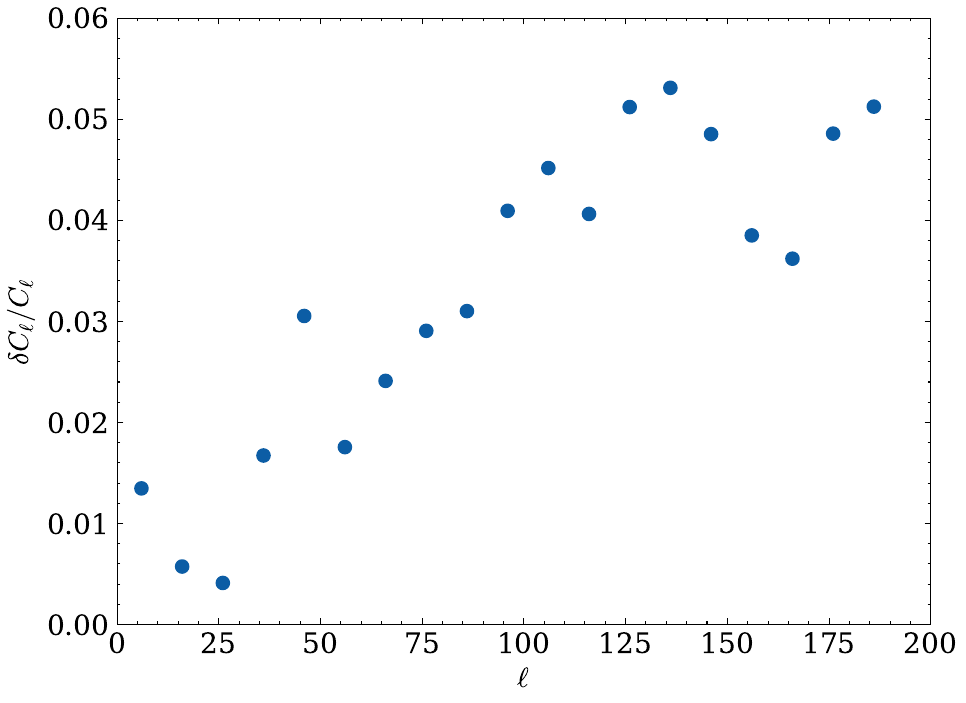}
    \caption{The impact of gain variation and noise. \textit{Left}: ratio of angular power spectrum with gain variation to the ideal case, before and after relative calibration. \textit{Right}: The angular band power for $\Delta \ell =10$ induced by the noise. }
    \label{fig:gain_and_noise}
\end{figure}        

Next we consider the impact of the gain variation and noise. In the left panel of Fig.~\ref{fig:gain_and_noise} we plot the ratio of the angular power spectrum with gain variation to that of the ideal case. The absolute calibration is performed to determine the initial phase. We plot both the case without relative calibration (blue curve, labelled as `before calibration') and the case with relative calibration (green curve, labelled as `after calibration'). We can see that in both cases the average of the angular power spectrum ratio are around 1. There are $10^{-2}$ level fluctuations in the angular power spectrum ratio for the case without relative calibration, showing that even without the relative calibration, the array could still make reasonably good measurement. This fluctuation is significantly reduced with the introduction of relative calibration. 

The addition of noise does not significantly change the angular power spectrum, even for the relatively large noise level of $N_{\rm day}=1$, i.e. just the data of one day. One can expect this error would be further reduced with integration. The angular power spectrum with noise has a shape similar to the reconstructed angular power spectrum in the ideal case, namely it is well recovered at $l\lesssim 200$, though slightly larger than the ideal case, which is as would be expected for a random noise power. 
We plot the band power $\delta C_l/C_l^{\rm recon}$ for bins of $\Delta \ell=10$ in the right panel of Fig.\ref{fig:gain_and_noise}, where $\delta C_l= C_l^{\rm +noise}-C_l^{\rm recon}$ is the power induced by the noise. Here we have included the effect of gain variation which has been corrected by the relative calibration. This band power is at a few percent level for a one day observation, but it can be reduced with longer integration time.

\subsection{Reconfiguration}

The analysis of the angular power spectrum transfer function in the last subsection indicates that a lack of long North-South baselines may be the reason for the error in reconstruction at $l \gtrsim 200$. The Tianlai pathfinder cylinder is 40 meters long, and the reason we have only used 12.4 m in the center section of the cylinder is due to the limited funding available, and the choice of one wavelength for the shortest spacing. As a result, the longest North-South baseline is much shorter than that of the East-West (30 m). To check whether this is indeed responsible for the abrupt drop of transfer function at $l \gtrsim 200$, we consider two cases for a different array configuration, where the maximum baseline length in the North-South is about 30 meters, the same as in the East-West:
\begin{itemize}
    \item[\textbullet] \textbf{Configuration 1}: we simply increase the spacing of feeds on each cylinder to $30 \divisionsymbol 12.4 \approx 2.42$ times the original spacing;
    \item[\textbullet] \textbf{Configuration 2}: we keep the position of central feeds on each cylinder unchanged, but move the 6 outer feeds on two ends further away, and the distance to the outermost feed of the central part follows a geometric progression $d_i = \Delta d^0 q^{i} \; (i = 1, 2, 3, 4, 5, 6)$, where $\Delta d^0$ is the current spacing between feeds, $q = (\frac{9 + 6  \Delta d^0}{ \Delta d^0})^{1/6}$.
\end{itemize}

\begin{figure}[htb]
    \centering
    \includegraphics[width=0.48\textwidth]{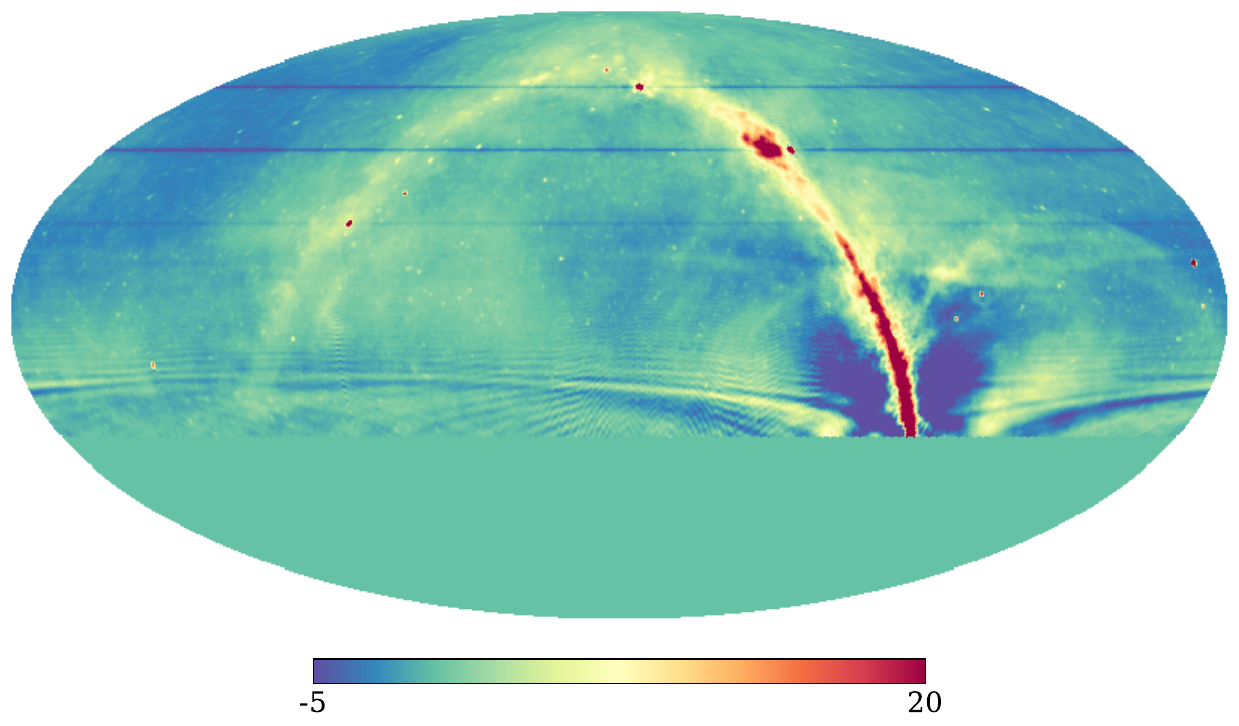}
    \includegraphics[width=0.48\textwidth]{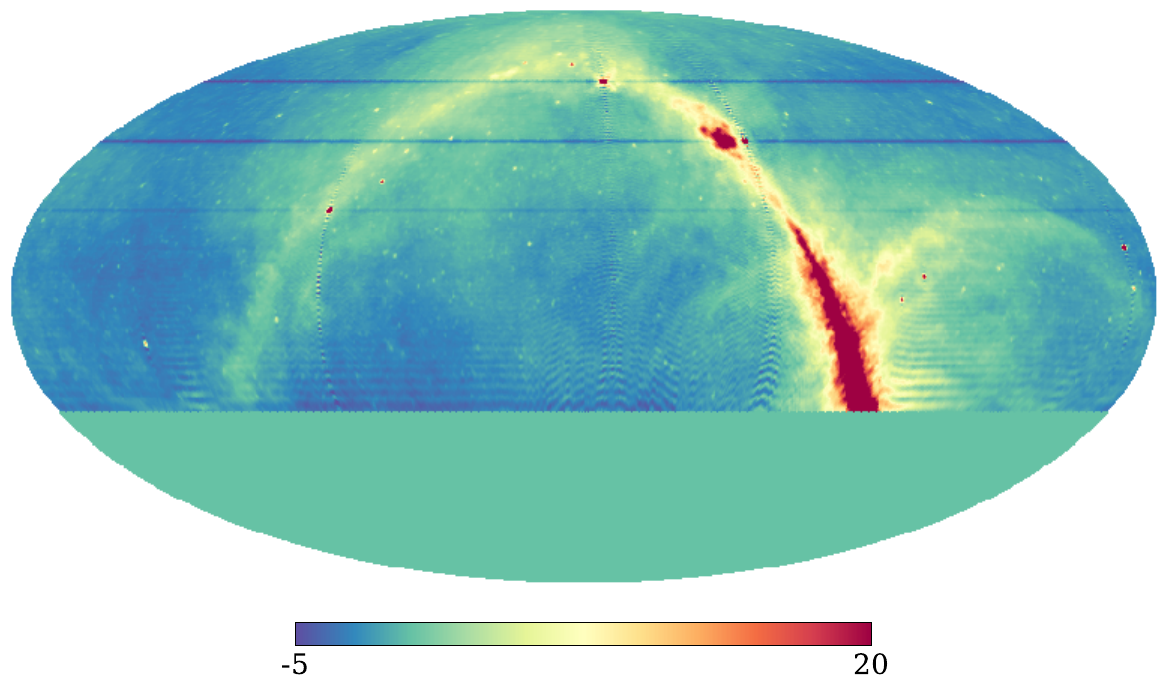}

    \caption{ 
    The reconstructed map with threshold $\epsilon=10^{-3}$ for the two expanding configurations. \textit{Left}: configuration 1; \textit{Right}: configuration 2. 
   }
\label{fig:reconstructed_map_ns_expanding}
\end{figure}

\begin{figure}[h!]
    \centering
\includegraphics[width=0.49\textwidth]{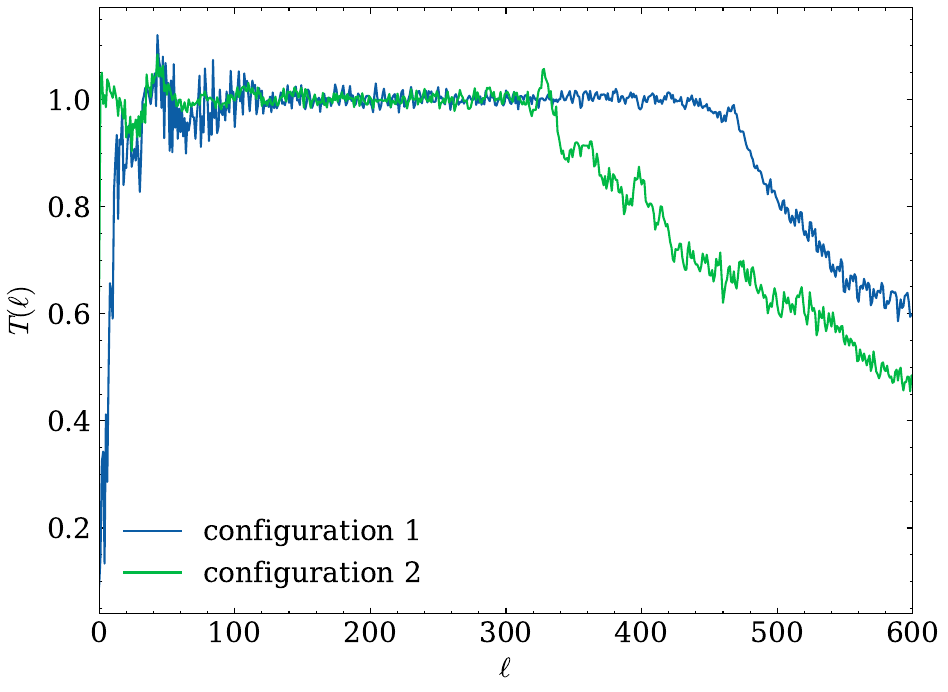}
\caption{
The transfer function  $C_{\ell}/C_{\ell}^{\rm input}$ with $\epsilon=10^{-3}$ for the two North-South expanding configurations.
}
    \label{fig:transfer_ns_expanding}
\end{figure}

In Fig. \ref{fig:reconstructed_map_ns_expanding}, we plot the reconstructed map with $\epsilon=10^{-3}$ for the two configurations in the ideal(noise-free) case, and in Fig.\ref{fig:transfer_ns_expanding} we plot the transfer function of angular power spectrum for the two cases with two thresholds.

For configuration 1, the comb-like structure in Fig.\ref{fig:reconstructed_map} disappeared, but the southern edge of the map is not well reconstructed, there is a deep blue region around the brightest part of the galactic plane, where it should be bright. We also see that the transfer function is much improved at $l \gtrsim 200$, and even for $\epsilon=10^{-3}$ it only starts to deviate from one at $\ \gtrsim 450$. However, there is a deep pit at $l<20$. This shows that with all North-South spacing increased by the same 2.42 factor, there is a dearth of short North-South baselines which probe the large scale variation in the North-South direction. It seems that the new defect in the reconstructed map is due to the missing of these short baselines.  

For configuration 2, where the maximum baseline along the North-South direction is equal to that of East-West, but some short baselines are preserved, the structure around the galactic plane is much better reconstructed. Even though we still adopted $\epsilon=10^{-3}$, comparing with Fig.\ref{fig:reconstructed_map} the comb-like artifact is significantly reduced.    
Compared with configuration 1, the deviation of transfer function for the configuration 2 starts at a lower $\ell$ value, but the $l<20$ modes are better reconstructed, as indicated by the fact that the transfer function remains near one at small $\ell$. So we conclude that at least for a single frequency, the array in the new configuration 2 has better performance than both the original configuration and configuration 1. 

\section{ Conclusion}
\label{sec:conclusion}

We have made a simulation for the observation of the Tianlai cylinder pathfinder array. This is done by first generating mock visibility from a {\tt cora} based sky model, and an unpolarized telescope beam model. The mock visibilities are then processed through the data processing pipeline {\tt tlpipe}, which has been used in the processing of the real data. In addition to the ideal case, we have also simulated random drifting of the complex gains of the array elements by Gaussian process, and the thermal noise according to the measured system Temperature. To correct for the variation of the complex gains, we also simulated the absolute and relative calibrations used in the Tianlai array. The absolute calibration is performed by making transit observation of the bright source Cyg A, which gives both the amplitude and phase of the complex gain. The relative calibration is performed by measuring a periodically broadcasting artificial calibrator noise signal to track the change of instrument phase. We have also added measurement noise for these measurements. 
Our simulation shows that the current absolute calibration algorithm produces an error at the $10^{-4}\sim 10^{-3}$ radian level. This error is due largely to the fact that the actual sky is not dominated by a single bright source, even when it is transiting.  The error induced by the thermal noise in the subsequent relative calibration is relatively small, only at the $10^{-5}$rad level. In order to further improve the measurement precision of the array, it is important to make improve the absolute calibration algorithm, probably by taking the additional sky source contribution into account. 

We then use the mock data to make maps of the sky by using the $m$-mode analysis, where the observed visibility data are decomposed into $m$-modes, and the spherical harmonics of the sky temperature fluctuations are solved from these $m$-modes. This solution has singularities due to incomplete $uv$ coverage. In the present work, we have used the Moore-Penrose pseudo-inverse method, to obtain the solution. In this method, modes with singular values smaller than a certain threshold are discarded. 
Due to the incomplete sampling of the $uv$ coverage, the reconstruction of the sky map is not perfect. There are sidelobes along the longitudinal lines. The $m=0$ modes which do not vary with time can not be recovered well from observation, and if we set these $m=0$ modes to zero, negative stripes along the latitude line passing through bright sources are generated. Furthermore, the Moore-Penrose pseudo-inverse algorithm may introduce some artifacts in the reconstructed map. While it is possible to suppress such artifact by adopting a smaller threshold parameter ($\epsilon < 10^{-5}$), in the presence of noise this is impractical.
We shall discuss other map-making algorithms elsewhere. In the present study, we focus on the map error induced by the calibration process and the noise. We find that the calibration error induced relatively small error in the map, which are most prominent near bright sources. The thermal noise is minimum in the part of the sky which is right above the telescope location, i.e. around declination of $\delta=44^\circ$, but increases toward north and south, where the sensitivity of the telescope degrades. 

We also investigated the angular power spectrum measurement. We find that the angular power spectrum is well recovered at $l\lesssim 200$. However, for $l> 200$, noise could introduce large error in the reconstructed map, and the angular power spectrum is systematically smaller for the higher modes. This is not surprising, as a baseline of length $b$ probes only modes within $l < 2\pi b/\lambda$, and the maximum length of the Tianlai cylinder pathfinder array baseline in the North-South direction is 12.4m, corresponding to $l \sim 195$. For higher $l$, there are modes which will not be sampled by the current Tianlai cylinder pathfinder array. We considered reconfigure the array, and find that an improvement can be achieved by moving six feeds on each end of the cylinder to increasingly larger spacings, to achieve longer baseline in the North-South direction, while preserving the shorter spacing in the center section of the cylinder. 
We also studied the impact of the thermal noise, and find that even in a single day observation it would be small compared with sky signal.

The present investigation has some limitations. We have only studied the case of a single frequency, and ignored polarization. We have also also limited the simulation to the current working algorithm of the Tianlai pipeline {\tt tlpipe}, which may not be optimal. It is conceivable that these algorithms could be further improved to reduce the numerical error. The array elements are treated as independent and identical. In fact, there are also cross-couplings between the different array elements, and each array element also has a slightly different response, which make the response of the telescope more complicated and introduce non-linearity. Such effects would be investigated in future studies.  

\begin{acknowledgements}
We thank Profs. Reza Ansari, Albert Stebbins and Peter Timbie for discussions. This work is supported by the National SKA Program of China, No.2022SKA0110100. Some of the simulation work is supported by the National Natural Science Foundation of China under Grant No. 1220030249.
\end{acknowledgements}

\bibliographystyle{raa}
\bibliography{main} 

\end{document}